\documentclass[final,twoside,11pt]{entics} 

\newif\ifignore % when set to true, additional text appears containing
\ignorefalse
\newcommand{\auxproof}[1]{
\ifignore\mbox{}\newline
\textbf{PROOF:} \dotfill\newline
{\it #1}\mbox{}\newline
\textbf{ENDPROOF}\dotfill
\fi}

\usepackage{enticsmacro}
\usepackage{graphicx}
\usepackage{etex} % to get more space
\usepackage{amsmath}
\usepackage{amssymb}
\usepackage{amstext}
\usepackage{mathtools}
\usepackage{mathrsfs}
\usepackage{stmaryrd}
\usepackage{xspace}
\usepackage{enumitem}
\usepackage{cmll}
\usepackage{url}
\usepackage{nicefrac}
\usepackage{pdfrender}

\usepackage{xypic}
\usepackage[all,cmtip]{xy}
\xyoption{tips}
\usepackage{tikzit}
\usetikzlibrary{circuits.ee.IEC}
\usetikzlibrary{shapes,shapes.geometric,shapes.misc}
\tikzstyle{white dot}=[inner sep=0mm, minimum size=1.5mm, draw=black, shape=circle, text depth=-0.2mm, draw=black, fill=white, tikzit category=nodes]
\tikzstyle{black dot}=[inner sep=0mm, minimum size=1.5mm, draw=black, shape=circle, draw=black, fill=black, tikzit category=nodes]
\tikzstyle{observed}=[inner sep=0mm, minimum size=5mm, draw=black, shape=circle, text depth=-0.2mm, draw=white, tikzit draw=gray, fill=white, tikzit category=dag]
\tikzstyle{latent}=[inner sep=0mm, minimum size=5mm, draw=black, shape=circle, text depth=-0.2mm, draw=black, fill=white, tikzit category=dag]
\tikzstyle{small box}=[shape=rectangle, text height=1.5ex, text depth=0.25ex, yshift=0.5mm, fill=white, draw=black, minimum height=6mm, yshift=-0.5mm, minimum width=6mm, font={\small}, tikzit category=boxes]
\tikzstyle{medium box}=[shape=rectangle, draw=black, fill=white, small box, minimum width=8mm, tikzit category=boxes]
\tikzstyle{semilarge box}=[shape=rectangle, draw=black, fill=white, small box, minimum width=12.5mm, tikzit category=boxes]
\tikzstyle{large box}=[shape=rectangle, draw=black, fill=white, small box, minimum width=15mm, tikzit category=boxes]
\tikzstyle{upground}=[circuit ee IEC, thick, ground, rotate=90, scale=1.5, inner sep=-2mm, tikzit shape=circle, tikzit fill=blue, tikzit category=points]
\tikzstyle{downground}=[circuit ee IEC, thick, ground, rotate=-90, scale=1.5, inner sep=-2mm, tikzit shape=circle, tikzit fill=green, tikzit category=points]
\tikzstyle{point}=[regular polygon, regular polygon sides=3, draw, scale=0.75, inner sep=-0.5pt, minimum width=9mm, fill=white, regular polygon rotate=180, tikzit category=points]
\tikzstyle{copoint}=[regular polygon, regular polygon sides=3, draw, scale=0.75, inner sep=-0.5pt, minimum width=9mm, fill=white, tikzit category=points]
\tikzstyle{uniform}=[point, fill=gray, tikzit shape=circle, scale=0.5]
\tikzstyle{label}=[font={\footnotesize}, text height=1.5ex, text depth=0.25ex, tikzit draw=blue, tikzit fill=white, tikzit category=labels]
\tikzstyle{left label}=[label, anchor=east, xshift=2mm, tikzit draw=green, tikzit fill=white, tikzit category=labels]
\tikzstyle{right label}=[label, anchor=west, xshift=-2mm, tikzit draw=purple, tikzit fill=white, tikzit category=labels]
\tikzstyle{disintegration}=[draw=black, fill={gray!50}, tikzit fill=gray, shape=rectangle, minimum width=1.6cm, minimum height=1.2cm, opacity=0.3]
\tikzstyle{empty diag}=[shape=rectangle, draw=darkgray, dashed, minimum width=8mm, minimum height=8mm, yshift=0.5mm]

\tikzstyle{diredge}=[->, >=latex]
\tikzstyle{dashed edge}=[-, dashed]

\ifdefined\mathsl\else
  \DeclareMathAlphabet{\mathsl}{\encodingdefault}{\rmdefault}{\mddefault}{\sldefault}
  \SetMathAlphabet{\mathsl}{bold}{\encodingdefault}{\rmdefault}{\bfdefault}{\sldefault}
\fi

\newenvironment{myproof}{\begin{trivlist} \item[\hskip \labelsep%
{\bf Proof.}]}{\end{trivlist}}

\newcommand{\QEDbox}{\square}
\newcommand{\QED}{\hspace*{\fill}$\QEDbox$}

\newcommand*{\fatten}[1][.4pt]{%
  \textpdfrender{
    TextRenderingMode=FillStroke,
    LineWidth={\dimexpr(#1)\relax},
  }%
}

\ifdefined\mathsl\else
  \DeclareMathAlphabet{\mathsl}{\encodingdefault}{\rmdefault}{\mddefault}{\sldefault}
  \SetMathAlphabet{\mathsl}{bold}{\encodingdefault}{\rmdefault}{\bfdefault}{\sldefault}
\fi

\makeatletter
\newcommand{\mathoverlap}[2]{\mathpalette\mathoverlap@{{#1}{#2}}}
\newcommand{\mathoverlap@}[2]{\mathoverlap@@{#1}#2}
\newcommand{\mathoverlap@@}[3]{\ooalign{$\m@th#1#2$\crcr\hidewidth$\m@th#1#3$\hidewidth}}
\makeatother

\newcommand{\klafter}{\mathbin{\mathoverlap{\circ}{\cdot}}}
\DeclareSymbolFont{T1op}{T1}{cmr}{m}{n}
\SetSymbolFont{T1op}{bold}{T1}{cmr}{bx}{n}
\DeclareMathSymbol{\mathguilsinglleft}{\mathopen}{T1op}{'016}
\DeclareMathSymbol{\mathguilsinglright}{\mathclose}{T1op}{'017}
\newcommand{\klin}[1]{\mathguilsinglleft#1\mathguilsinglright}

\newcommand{\idmap}[1][]{\ensuremath{\mathrm{id}_{#1}}}
\newcommand{\after}{\mathrel{\circ}}
\newcommand{\pull}{\mathrel{\mathchoice%
   {\scalebox{-0.5}[1]{$\gg=$}}
   {\scalebox{-0.5}[1]{$\gg{\kern-1.5ex}=$}}
   {\scalebox{-0.5}[1]{${\kern.5ex}\scriptstyle\gg{\kern-0.2ex}={\kern.5ex}$}}
   {\scalebox{-0.5}[1]{$\scriptscriptstyle\gg=$}}}}
\newcommand{\push}{\mathrel{\mathchoice%
   {\scalebox{-0.5}[1]{$=\ll$}}
   {\scalebox{-0.5}[1]{$={\kern-1.5ex}\ll$}}
   {\scalebox{-0.5}[1]{${\kern.5ex}\scriptstyle={\kern-0.2ex}\ll{\kern.5ex}$}}
   {\scalebox{-0.5}[1]{$\scriptscriptstyle=\ll$}}}}

\newcommand{\partdrawadd}{\ensuremath{\mathsl{PDA}}}
\newcommand{\setdrawadd}{\ensuremath{\mathsl{SDA}}}
\newcommand{\partdrawdelete}{\ensuremath{\mathsl{PDD}}}

\newcommand{\multinomial}[1][]{\ensuremath{\mathsl{mn}[#1]}}
\newcommand{\partmultinomial}{\ensuremath{\mathsl{pamn}}}
\newcommand{\swapmultinomial}[1][]{\ensuremath{\mathsl{swmn}[#1]}}

\newcommand{\stirling}[1][]{\ensuremath{\mathsl{stir}[#1]}}
\newcommand{\poissonname}{\mathsl{pois}}
\newcommand{\poisson}[1][]{\ensuremath{\poissonname[#1]}}
\newcommand{\ewens}[1][]{\ensuremath{\mathsl{ew}[#1]}}

\newcommand{\setin}[3]{\{#1\in#2\;|\;#3\}}
\newcommand{\supp}{\mathsl{supp}}
\newcommand{\acc}{\mathsl{acc}}
\newcommand{\arr}{\mathsl{arr}}
\newcommand{\tupperm}{\mathsl{tp}}
\newcommand{\eltperm}{\mathsl{ep}}
\newcommand{\mulcount}{\mathsl{mc}}
\newcommand{\stack}{\mathsl{stk}}
\newcommand{\iid}{\mathsl{iid}}
  
\newcommand{\size}{\ensuremath{\mathsl{size}}}  
\newcommand{\som}{\ensuremath{\mathsl{sum}}}  
\newcommand{\maal}{\ensuremath{\mathsl{prod}}}  
\newcommand{\Perm}{\ensuremath{\mathsl{Perm}}} % \wp
\newcommand{\MP}{\ensuremath{\mathsl{MP}}} % \wp
\newcommand{\Div}{\ensuremath{\mathfrak{D}{\kern-1.0pt}V}}
\newcommand{\facto}[1]{\ensuremath{#1{\kern-2.5pt}\raisebox{-2.5pt}{\includegraphics[width=0.9em]{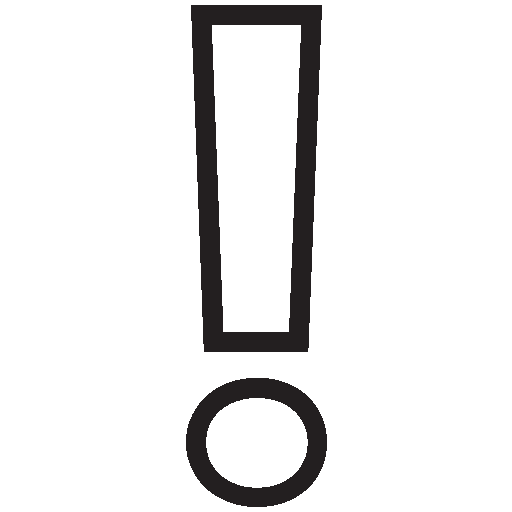}}}}
\newcommand{\sfacto}[1]{\ensuremath{#1{\kern-1.5pt}\raisebox{-1.5pt}{\includegraphics[width=0.6em]{exclamation}}}}
\newcommand{\coefm}[1]{\ensuremath{\fatten[0.6pt]{(}{\kern1pt}#1{\kern1pt}\fatten[0.6pt]{)}}}
\newcommand{\partcoefm}[1]{\coefm{#1}_{P}}

\newcommand{\setsize}[1]{|{\kern.1em}#1{\kern.1em}|}
\newcommand{\bigsetsize}[1]{\big|{\kern.1em}#1{\kern.1em}\big|}
\newcommand{\tuple}[1]{\langle#1\rangle}

\newcommand{\ket}[1]{\ensuremath{|{\kern.1em}#1{\kern.1em}\rangle}}
\newcommand{\bigket}[1]{\ensuremath{\big|{\kern.1em}#1{\kern.1em}\big\rangle}}
\newcommand{\ketstrut}{\vrule height 10pt depth 5pt width 0pt}
\newcommand{\Bigket}[1]{\ensuremath{\left|\ketstrut{\kern.1em}#1{\kern.005em}\right>}}

\newcommand{\andthen}{\mathrel{\&}}

\newcommand{\incr}[2]{#1\!\mathrel{\mathrm{+{\kern-.1em}+}}\!#2}

\newcommand{\distributionsymbol}{\mathcal{D}}

\newcommand{\multisetsymbol}{\mathcal{M}}

\newcommand{\Dst}{\distributionsymbol}
\newcommand{\infDst}{\ensuremath{\mathcal{D}_{\infty}}}
\newcommand{\Mlt}{\multisetsymbol}

\newcommand{\UF}{\ensuremath{\mathcal{U}{\kern-.75ex}\mathcal{F}}}

\newcommand{\Kl}{\mathcal{K}{\kern-.4ex}\ell}

\newcommand{\EM}{\mathcal{E}{\kern-.4ex}\mathcal{M}}

\newcommand{\NNO}{\mathbb{N}}
\newcommand{\pNNO}{\mathbb{N}_{>0}}

\newcommand{\R}{\mathbb{R}}

\newcommand{\pR}{\R_{> 0}}

\newcommand{\Pred}{\ensuremath{\mathrm{Pred}}}

\newcommand{\Ef}{\ensuremath{\mathcal{E}{\kern-.5ex}f}}
\newcommand{\intd}{{\kern.2em}\mathrm{d}{\kern.03em}}
\newcommand{\indic}[1]{\mathbf{1}_{#1}}

\newcommand{\OF}{\ensuremath{\mathcal{O}{\kern-.1em}\mathcal{F}}}
\newcommand{\Closed}{\ensuremath{\mathcal{C}{\kern-.05em}\ell}}

\newcommand{\congrightarrow}{\mathrel{\smash{\stackrel{
           \raisebox{.5ex}{$\scriptstyle\cong$}}{
           \raisebox{0ex}[0ex][0ex]{$\rightarrow$}}}}}

\newsavebox\sbpto
\savebox\sbpto{\begin{tikzpicture}[baseline=-2.4pt]
            \filldraw[fill=white,draw=white] circle (1.4pt);
            \filldraw[fill=white,draw=black,line width=0.2pt] circle (2.0pt);
                \end{tikzpicture}}
\newcommand\chanto{\mathrel{\ooalign{$\rightarrow$\cr
            \hfil\!$\usebox\sbpto$\hfil\cr}}}

\newsavebox\sbground
\savebox\sbground{\begin{tikzpicture}[circuit ee IEC,yscale=1,xscale=1]
                \draw (0,-2ex) to (0,0ex) node[ground,rotate=90,xshift=.65ex] {};
                \end{tikzpicture}}
\newcommand\ground{\mathbin{\text{\raisebox{0.2ex}{\usebox\sbground}}}}
\newsavebox\sbunif
\savebox\sbunif{\begin{tikzpicture}[circuit ee IEC,yscale=1,xscale=1]
                \draw (0,0) to (0,2ex) node[ground,rotate=270,xshift=2.5ex] {};
                \end{tikzpicture}}

\tikzset{dot/.style =
  {inner sep=0mm,minimum width=1mm,minimum height=1mm,
    draw,shape=circle}}
\tikzset{minicopy/.style = {dot,fill,text depth=-0.2mm}}
\newsavebox\sbcopier
\savebox\sbcopier{%
  \begin{tikzpicture}[baseline=0pt]
    \node[minicopy,scale=.7] (a) at (0,3.6pt) {};
    \draw (a) -- +(-90:.30);
    \draw (a) -- +(45:.35);
    \draw (a) -- +(135:.35);
  \end{tikzpicture}}
\newcommand{\minicopy}{\mathord{\usebox\sbcopier}}

\makeatother
 \newdir{ >}{{}*!/-7.5pt/@{>}}
 \newdir{|>}{!/4.5pt/@{|}*:(1,-.2)@^{>}*:(1,+.2)@_{>}}
 \newdir{ |>}{{}*!/-3pt/@{|}*!/-7.5pt/:(1,-.2)@^{>}*!/-7.5pt/:(1,+.2)@_{>}}

\def\conf{MFPS 2022} 	%%Fill in the acronym for your conference (with year)
\volume{1}			%Fill in the ENTICS volume number here
\def\volu{1}			% and here
\def\papno{9}			%Fill in your paper number here
\def\mydoi{{\normalshape \href{https://doi.org/10.46298/entics.10520}{10.46298/entics.10520}}}
\def\copynames{B. Jacobs} %% Fill in the first initial and last name of the authors
\def\runtitle{Sufficient Statistics...}

\def\lastname{Jacobs}

\begin{document}

\begin{frontmatter}

\title{Sufficient Statistics and Split Idempotents \\
    in Discrete Probability Theory}

\author{Bart Jacobs\thanksref{ALL}\thanksref{myemail}}

\address{Institute for Computing and Information Sciences (iCIS) 
\\ 
Radboud University
\\
Nijmegen, The Netherlands.}  							

\thanks[ALL]{Thanks are due to Dario Stein and Dusko Pavlovic for
  lively discussions.  }

\thanks[myemail]{Email: \href{mailto:bart@cs.ru.nl} {\texttt{\normalshape
        bart@cs.ru.nl}}} 

%\def\titlerunning{Sufficient Statistic}
%\def\authorrunning{B. Jacobs}

%\date{\small \today}
%\date{}

\begin{abstract} 
A sufficient statistic is a deterministic function that captures an
essential property of a probabilistic function (channel,
kernel). Being a sufficient statistic can be expressed nicely in terms
of string diagrams, as Tobias Fritz showed recently, in adjoint
form. This reformulation highlights the role of split idempotents, in
the Fisher-Neyman factorisation theorem. Examples of a sufficient
statistic occur in the literature, but mostly in continuous
probability. This paper demonstrates that there are also several
fundamental examples of a sufficient statistic in discrete
probability. They emerge after some combinatorial groundwork that
reveals the relevant dagger split idempotents and shows that a
sufficient statistic is a deterministic dagger epi.
\end{abstract}

\begin{keyword}
Discrete probability theory, sufficient statistic, disintegration,
string diagram, split idempotent
\end{keyword}

\end{frontmatter}

\section{Introduction}\label{IntroSec}

The notion of a \emph{sufficient statistic} plays an important role in
statistics, but a precise definition is hard to find. Informally, it
involves a function $s\colon X \rightarrow Y$ that expresses a
characteristic property about certain parameterised probability
distributions $p(\theta)$ on $X$. The statistic provides enough
information such that this distribution can be reconstructed from the
push-forward of $p(\theta)$ along $s$. The original formulation has
been introduced and studied in the 1920s by Ronald Fisher,
see~\cite{Fisher22}.

A precise formulation has been given recently by Tobias Fritz as part
of his efforts to express basic concepts and results from probability
theory and statistics categorically, see~\cite{Fritz20} and
also~\cite{FritzGPR20,FritzR20}. This approach uses \emph{Markov
categories}, which are symmetric monoidal categories, with a final
object as tensor unit, and with copiers (studied as `CD-categories'
in~\cite{ChoJ19}). String diagrams provide a powerful and intuitive
language for these Markov categories.

Within this setting, the notion of sufficient statistic is formalised
as an equation of string diagrams, see~\cite[Defn.~14.3]{Fritz20}. It
can be read as a (symmetric) adjoint property, more in the style of
adjoint matrices then adjoint functors. One of the highlights
in~\cite[Thm.~14.5]{Fritz20} is an abstract formalisation (and proof)
of the Fisher-Neyman factorisation theorem (see also
\textit{e.g.}~\cite[Prop~4.10]{BernardoS00}
or~\cite[\S3.3]{SuhovK05}). It gives a necessary and sufficient
condition for the existence of a sufficient statistic, in terms of a
split idempotent. Thus, \cite{Fritz20} provides a great clarification
of what a sufficient statistic is all about, via abstraction. But it
does not really deal with examples of sufficient statistics.

The existing descriptions of sufficient statistics in the literature
have several disadvantages.
\begin{itemize}
\item The examples are mostly in continuous probability theory. There
  are hardly any illustrations in discrete probability theory, except
  the one with sums and Poisson distributions that we reproduce in
  Section~\ref{SumSuffStatSec}.  This is a pity, because there are
  several fundamental discrete instances.

\item Typically in these examples, only the sufficiency condition of
  the Fisher-Neyman factorisation theorem is established ---
  consisting of the disappearance of the parameter after
  conditioning. These descriptions stop short of giving the full
  picture, with the relevant split idempotent that does the real work
  and produces the adjoint situation.

\item These examples do not match the abstract description
  of~\cite{Fritz20} --- understandably so, given that~\cite{Fritz20}
  is a very recent publication.
\end{itemize}

\noindent Thus there is room for a fresh look at sufficiency of
statistics, given the clear and abstract reformulation provided
recently by~\cite{Fritz20}. That is what the current paper will
do. The emphasis is on discrete probability theory, since it already
offers ample material. This restriction has the (technical) benefit
that we do not have to bother with ``almost surely equal'', as in
continuous probability. Thus, the paper only looks at one particular
kind of Markov category, namely the Kleisli category of the discrete
distribution monad.

This paper elaborates examples, but it does not offer new (category)
theory. We like to compare examples of sufficient statistics with
examples of adjunctions, like: the forgetful functor from compact
Hausdorff spaces to sets has a left adjoint given by
ultrafilters. Proving this is quite a bit of work. It offers valuable
mathematical insight.

Similarly, we claim that the examples of sufficient statistics that we
describe below offer valuable insight in the relevant mathematical
structures. As is often the case in discrete probability theory, the
nature of these structures is combinatorial. Indeed, this paper
develops some new combinatorial results, especially about (multiset)
partitions, see \textit{e.g.}~Proposition~\ref{MulcountProp} below
(extending~\cite{Jacobs22c}). The emphasis is on uncovering the
relevant split idempotents.

We recall that a split idempotent is a map $f\colon A \rightarrow A$
that can be written as $f = s \after r$, where $r \after s = \idmap$.
In such a situation, the map $r$ is called a retraction and $s$ a
section.  It is easy to see that the section $s$ is the equaliser of
$f,\idmap \colon A \rightrightarrows A$, and that the retraction $r$
is their coequaliser. Such a splitting of $f$, if it exists is unique,
up-to-isomorphism. Split idempotents are used in the Karoubi envelope,
the idempotent-splitting completion of a category. Splittings of
dagger (self-adjoint) idempotents form classical objects in a quantum
setting~\cite{Selinger08}. Here we also have daggers, relative to a
prior distribution. We show that a sufficient statistic (in discrete
probability) is a deterministic dagger epi (see~\cite{HeunenJ10a}), as
retraction part of a split dagger idempotent, see
Lemma~\ref{DetDaggerLem}.

\auxproof{
Let $r \colon A \rightarrow B$ and $s\colon B\rightarrow A$.
\begin{itemize}
\item \textbf{The section $s$ is equaliser}: $f \after s = s \after r
  \after s = s = \idmap \after s$.  Moreover, if $f\after g = g =
  \idmap \after g \colon C \rightarrow A$, then there is $\overline{g}
  \coloneqq r \after g \colon C \rightarrow B$ satisfying:
\begin{itemize}
\item $s \after \overline{g} = s \after r \after g = f \after g = g$.

\item if also $s \after h = g$, then $h = r \after s \after h = r
  \after g = \overline{g}$.
\end{itemize}

\item \textbf{The retraction $r$ is coequaliser}: $r \after f = r
  \after s \after r = r = r \after \idmap$. If $g \after f = g = g
  \after \idmap\colon A \rightarrow C$, then $\overline{g} \coloneqq =
  g \after s \colon B \rightarrow C$ satisfies:
\begin{itemize}
\item $\overline{g} \after r = g \after s \after r = g \after f = g$.

\item if also $h \after r = g$, then $h = h \after r \after s = g
  \after s = \overline{g}$.
\end{itemize}
\end{itemize}
}

The paper starts with a relatively long section on background
material, covering multisets, discrete probability distributions,
channels and their daggers, updating, and partitions. These partitions
are special multisets over the natural numbers that have been studied
in their own right~\cite{Andrews98}, but also to capture mutations in
population biology (see
\textit{e.g.}~\cite{Crane16,Ewens72,Joyce98,Kingman78a,Kingman78b,Tavare21};
we offer some new result. Section~\ref{SuffStatSec} repeats the
description of sufficient statistics from~\cite{Fritz20} and puts it
in perspective, in relation to disintegration. The role of split
idempotents is emphasised to obtain examples of sufficient statistics.
The subsequent four sections~\ref{AccSuffStatSec} --
\ref{SumSuffStatSec} cover particular illustrations of the notion of
sufficient statistic. Section~\ref{AccSuffStatSec} looks at
accumulation (from sequences to multisets) as sufficient statistic for
independent and identically distributed (iid) elements in sequences.
This sufficiency situation, in Diagram~\ref{AccSuffStatEqn}, captures
the very basic relations at the heart of discrete probability theory.
Section~\ref{MulcountSuffStatSec} introduces a new sufficiency
situation for partitions, building on the multiplicity count function
(from~\cite{Jacobs22c}).  Sections~\ref{SizeSuffStatSec}
and~\ref{SumSuffStatSec} review two examples from the literature in
the current setting, with the adjoint description based on split
idempotents.

\section{Background}\label{BackgroundSec}

We briefly review the essence of multisets, distributions and
partitions, and also of channels as probabilistic functions.
The main goal is to fix notation.

\subsection{Multisets}\label{MultisetSubsec}

A multiset (or bag) is a finite `subset' in which elements may occur
multiple times. There are two equivalent representations for a
multiset with elements from a set $X$.
\begin{itemize}
\item As formal finite sums $\sum_{i}n_{i}\ket{x_i}$ of elements
  $x_{i}\in X$ with multiplicity $n_{i}\in\NNO$.

\item As functions $\varphi\colon X \rightarrow \NNO$ with finite
  support $\supp(\varphi) \coloneqq \setin{x}{X}{\varphi(x) > 0}$.
\end{itemize}

\noindent We switch between these representations whenever convenient.

We write $\Mlt(X)$ for the set of multisets over $X$. It is the free
commutative monoid on $X$. This $\Mlt$ forms a monad on the category
of sets. We only mention how functoriality works: for a function
$f\colon X \rightarrow Y$ there is a function $\Mlt(f) \colon \Mlt(X)
\rightarrow \Mlt(Y)$ defined by:
\begin{equation}
\label{MltFunctorEqn}
\begin{array}{rclcrcl}
\Mlt(f)\Big(\sum_{i}n_{i}\ket{x_i}\Big)
& \coloneqq &
\sum_{i} n_{i}\bigket{f(x_{i})}
& \quad\mbox{or, equivalently,}\quad &
\Mlt(f)(\varphi)
& = &
\displaystyle\sum_{y\in Y} \left(\sum_{x\in f^{-1}(y)} \varphi(x)\right)\bigket{y}.
\end{array}
\end{equation}

\noindent With a multiset $\varphi$ we associate the following numbers.
\begin{itemize}
\item $\|\varphi\| \coloneqq \sum_{x}\varphi(x)$ is the size of
  $\varphi$: the total number of elements in $\varphi$, including
  multiplicities;

\item $\facto{\varphi} \coloneqq \prod_{x}\varphi(x)!$ is the product
  of the factorials of multiplicities;

\item $\coefm{\varphi} \coloneqq \frac{\|\varphi\|!}{\sfacto{\varphi}}
  = \frac{(\sum_{x}\varphi(x))!}{\prod_{x} \varphi(x)!}$ is the
  multinomial coefficient;

\item $\binom{n}{\varphi} \coloneqq \frac{n!}{\sfacto{\varphi}\cdot
  (n-\|\varphi\|)}$ when $n\geq \|\varphi\|$.
\end{itemize}

\noindent For a number $K$ we define the subset $\Mlt[K](X) \subseteq
\Mlt(X)$ of multisets of size $K$:
\[ \begin{array}{rcl}
\Mlt[K](X)
& \coloneqq &
\setin{\varphi}{\Mlt(X)}{\|\varphi\| = K}.
\end{array} \]

\noindent There is an `accumulation' function $\acc \colon X^{K}
\rightarrow \Mlt[K](X)$ that accumulates multiple occurrences in a
sequence: $\acc(x_{1}, \ldots, x_{K}) = 1\ket{x_{1}} + \cdots +
1\ket{x_{K}}$. For instance $\acc(a,a,b,a,c,b) = 3\ket{a} + 2\ket{b} +
1\ket{c}$.

For a finite set $A$ we write $\setsize{A}\in\NNO$ for its number of
elements and $\Perm(A)$ for the set of permutations $\pi \colon A
\congrightarrow A$. Recall that $\bigsetsize{\Perm(A)} =
\setsize{A}!$.

\begin{lemma}
\label{AccLem}
\begin{enumerate}
\item \label{AccLemSize} For a multiset $\varphi\in\Mlt[K](X)$ there
  are $\coefm{\varphi}$ many sequences $\ell\in X^{K}$ with $\acc(\ell) =
  \varphi$, that is, $\bigsetsize{\acc^{-1}(\varphi)} =
  \coefm{\varphi}$.

\item \label{AccLemSum} If $X$ is finite, then
  $\displaystyle\sum_{\varphi\in\Mlt[K](X)} \coefm{\varphi} =
  \setsize{X}^{K}$. \QED
\end{enumerate}
\end{lemma}

\subsection{Distributions}\label{DistributionSubsec}

A (discrete finite probability) distribution over a set $X$ is a
formal sum $\sum_{i} r_{i}\ket{x_i}$, where $x_{i}\in X$ and $r_{i}\in
[0,1]$ with $\sum_{i}r_{i} = 1$. Alternatively, it is a function
$\omega \colon X \rightarrow [0,1]$ with finite support and
$\sum_{x}\omega(x) = 1$. We write $\Dst(X)$ for the set of
distributions over $X$. Such distributions may also be called states,
see~\cite{Jacobs17b}. This $\Dst$ is a monad too, like
$\Mlt$. Functoriality works as in~\eqref{MltFunctorEqn}.

For two distributions $\omega\in\Dst(X)$ and $\rho\in\Dst(Y)$ one can
form a product distribution $\omega\otimes\rho\in\Dst(X\times Y)$ via
$(\omega\otimes\rho)(x,y) = \omega(x)\cdot\rho(y)$. Equivalently,
$\omega\otimes\rho = \sum_{x,y} \omega(x)\cdot\rho(y)\bigket{x,y}$.
For a single distribution $\omega\in\Dst(X)$ we write the $K$-fold
(tensor) product as:
\begin{equation}
\label{IIDEqn}
\begin{array}{rcccl}
\iid[K](\omega)
& \coloneqq &
\omega^{K}
& = &
\omega\otimes\cdots\otimes\omega \;\in\; \Dst\big(X^{K}\big).
\end{array}
\end{equation}

\noindent The abbreviation `iid' stands for `independent and
identically distributed'. Marginalisation of a `joint' state $\tau\in
\Dst(X_{1}\times\cdots\times X_{n})$ is the operation of projecting
$\tau$ to a state in $\Dst(X_{i})$, on one of the components, obtained
via functoriality, as $\Dst(\pi_{i})(\tau)$.

We shall make frequent use of multinomial distributions, which we
describe via a function $\multinomial[K] \colon \Dst(X) \rightarrow
\Dst\big(\Mlt[K](X)\big)$. We think of $X$ as a set of colours and of
$\omega\in\Dst(X)$ as an urn with coloured balls, where $\omega(x)\in
        [0,1]$ is the probability of drawing a ball of colour $x\in
        X$. Then, $\multinomial[K](\omega) \in
        \Dst\big(\Mlt[K](X)\big)$ is a distribution on multisets of
        size $K$, corresponding to draws of $K$-many balls from the
        urn (with replacement). This distribution is given by:
\begin{equation}
\label{MultinomialEqn}
\begin{array}{rcccl}
\multinomial[K](\omega)
& \coloneqq &
\Dst\big(\acc\big)\Big(\iid[K](\omega)\Big)
& = &
\displaystyle\sum_{\varphi\in\Mlt[K](X)} \coefm{\varphi} \cdot
   \prod_{x\in X} \omega(x)^{\varphi(x)} \,\bigket{\varphi}.
\end{array}
\end{equation}

\noindent See~\cite{Jacobs21b,Jacobs22a} for more details. As
illustration, we consider draws of size~$3$ with a distribution
$\frac{1}{8}\ket{a} + \frac{1}{2}\ket{b} + \frac{3}{8}\ket{c}$,
corresponing to an urn with $8$ balls, one of colour $a$, 4 of colour
$b$, 3 of colour $c$. These draws (of size $3$) then come with the
following distribution.
\[ \begin{array}{rcl}
\lefteqn{\textstyle\multinomial[3]\Big(\frac{1}{8}\ket{a} + 
   \frac{1}{2}\ket{b} + \frac{3}{8}\ket{c}\Big)}
\\[+0.6em]
& = &
\frac{1}{512}\Bigket{3\ket{a}} + 
   \frac{3}{128}\Bigket{2\ket{a} + 1\ket{b}} + 
   \frac{3}{32}\Bigket{1\ket{a} + 2\ket{b}} + 
   \frac{1}{8}\Bigket{3\ket{b}} + 
   \frac{9}{512}\Bigket{2\ket{a} + 1\ket{c}} \, + 
\\[+0.6em]
& & \;\; 
   \frac{9}{64}\Bigket{1\ket{a} + 1\ket{b} + 1\ket{c}} + 
   \frac{9}{32}\Bigket{2\ket{b} + 1\ket{c}} + 
   \frac{27}{512}\Bigket{1\ket{a} + 2\ket{c}} + 
   \frac{27}{128}\Bigket{1\ket{b} + 2\ket{c}} + 
   \frac{27}{512}\Bigket{3\ket{c}}
\end{array} \]

% S = Space('a', 'b', 'c')
% print( Multinomial(3)(DState([Frac(1,8), Frac(1,2), Frac(3,8)], S)) )

In the previous subsection we have seen the accumulation
\emph{function} $\acc \colon X^{K} \rightarrow \Mlt[K](X)$. In the
other direction there is a \emph{channel} $\arr \colon \Mlt[K](X)
\chanto X^{K}$, called arrangement (see~\cite{Jacobs21b}); it is given
by:
\begin{equation}
\label{ArrEqn}
\begin{array}{rcl}
\arr(\varphi)
& \coloneqq &
\displaystyle\sum_{\vec{x} \in \acc^{-1}(\varphi)} \frac{1}{\coefm{\varphi}} \,
   \bigket{\vec{x}}.
\end{array}
\end{equation}

\noindent This uniform distribution is well-defined, by
Lemma~\ref{AccLem}~\eqref{AccLemSize}. We shall see that accumulation
and arrangement form a split idempotent, for a special `Kleisli'
composition, see Lemma~\ref{ArrLem}.

We introduced distributions $\omega\in\Dst(X)$ with \emph{finite}
support. In Section~\ref{SumSuffStatSec} we drop this finiteness
requirement and shall work with more general discrete probability
distributions. We then write $\infDst(X)$ for the set of such
distributions.

\subsection{Channels}\label{ChannelSubsec}

We have mentioned that taking distributions $\Dst$ is a monad. The
associated `Kleisli' maps $X \rightarrow \Dst(Y)$ are called
\emph{channels} and will be written as $X \chanto Y$, with a circle on
the shaft. These arrows $\chanto$ are the morphisms in the (symmetric
monoidal) Kleisli category $\Kl(\Dst)$ of the distribution monad
$\Dst$. We recall the basic operations, of Kleisli extension $\push$
and Kleisli composition $\klafter$.

Given a channel $c\colon X \chanto Y$ and a distribution / state
$\omega\in\Dst(X)$ we write $c \push \omega \in \Dst(Y)$ for the
\emph{state transformation} along $c$, obtained via Kleisli extension:
\[ \begin{array}{rcl}
\big(c \push \omega\big)(y)
& \coloneqq &
\displaystyle\sum_{x\in X} \omega(x) \cdot c(x)(y).
\end{array} \]

\noindent Via $\push$ we can define (Kleisli) composition $\klafter$
of channels as $\big(d \klafter c\big)(x) = d \push c(x)$. The unit
$\eta \colon X \rightarrow \Dst(X)$ of the monad $\Dst$, with $\eta(x)
= 1\ket{x}$, is the identity (channel) for this composition
$\klafter$.

For two channels $c\colon X \chanto A$ and $d\colon Y \chanto B$ there
is a parallel product $c \otimes d \colon X\times Y \chanto A\times B$
defined as pointwise tensor: $\big(c \otimes d)(x,y) = c(x) \otimes
d(y)$. When the domains $X,Y$ are equal we can form a tuple channel
$\tuple{c,d} \coloneqq (c\otimes d) \after \Delta \colon X \rightarrow
A\times B$, where $\Delta \colon X \rightarrow X\times X$ is the copy
function $\Delta(x) = (x,x)$. We shall see special instances of this
tupling of channels when $c$ or $d$ is the identity channel. We shall
write such tuples as $\tuple{c,\idmap}$ or $\tuple{\idmap,d}$. We
spell out the associated state transformation:
\begin{equation}
\label{TupleKetEqn}
\begin{array}{rcl}
\tuple{c,\idmap} \push \omega
& = &
\displaystyle\sum_{x\in X, y\in Y} \omega(x) \cdot c(x)(y) \,\bigket{x,y}.
\end{array}
\end{equation}

Each function $f\colon X \rightarrow Y$ gives rise to a
\emph{deterministic} channel $\klin{f} \coloneqq \eta \after f \colon
X \chanto Y$, via the unit $\eta$ of the monad, so that $\klin{f}(x) =
1\bigket{f(x)}$. These deterministic channels, of the form $\klin{f}$,
can be characterised as those channels that commute with copyiers
$\Delta$. They satisfy: $\klin{g} \klafter \klin{f} = \klin{g \after
  f}$ and $d \klafter \klin{f} = d \after f$ and $\klin{f} \klafter c
= \Dst(f) \after c$. Often we omit the brackets $\klin{\mbox{-}}$ when
it is implicitly clear that an ordinary function is promoted to a
channel. We do so in particular for projection and copy functions
$\pi_i$ and $\Delta$, and also for accumulation in the next result.

\begin{lemma}
\label{ArrLem}
Accumulation and arrangement~\eqref{ArrEqn} form (the section and
retraction of) a split idempotent in the Kleisli category $\Kl(\Dst)$
of the distribution monad $\Dst$, since: $\acc \klafter \arr =
\idmap$. The resulting split idempotent channel $\tupperm \coloneqq
\arr \klafter \acc \colon X^{K} \chanto X^{K}$ is tuple-permutation
(``transposition''):
\[ \begin{array}{rcl}
\tupperm\big(x_{1}, \ldots, x_{K})
& = &
\displaystyle\sum_{\pi\in\Perm(\{1,\ldots, K\})} \frac{1}{K!}
   \,\Bigket{x_{\pi(1)}, \ldots, x_{\pi(K)}}.
\end{array} \eqno{\QEDbox} \]
\end{lemma}

For convenience and clarity, we shall use the graphical notation of
string diagrams for channels, as probabilistic computations, where
information is flowing upwards. A channel is represented as a box, and
wires are typed by the input and output sets of these channels. We
assume a basic level of familiarity with these string diagrams and
refer to~\cite{ChoJ19,Fritz20,Stein22} for more information about
their use in probability theory. In this paper these string diagram
are interpreted in the Kleisli category $\Kl(\Dst)$, or
$\Kl(\infDst)$, but not in general Markov categories.

\subsection{Predicates, updating and daggers}\label{PredicateSubsec}

A predicate on a set $X$ is a function $p\colon X \rightarrow [0,1]$.
We shall write $\Pred(X)$ for the set of all such (fuzzy) predicates
on $X$. Updating (or conditioning) of a state/distribution is a basic
operation in probabilistic reasoning, where $\omega\in\Dst(X)$ is
updated in the light of evidence given by a predicate $p\in\Pred(X)$.
We first write $\omega\models p \coloneqq \sum_{x} \omega(x)\cdot
p(x)$ for the validity (expected value) of $p$ in $\omega$. If this
validity is non-zero, we can define the updated distribution
$\omega|_{p} \in\Dst(X)$ as the normalised product:
\[ \begin{array}{rcl}
\omega|_{p}(x)
& \coloneqq &
\displaystyle\frac{\omega(x)\cdot p(x)}{\omega\models p}.
\end{array} \]

\noindent One can then prove that updating makes the predicate $p$
`more true', in the sense that $\omega|_{p} \models p \,\geq\,
\omega\models p$, see \textit{e.g.}~\cite{Jacobs19c,Jacobs21b} for
details.

Along a channel $c\colon X \chanto Y$ one can do backwards
predicate transformation: for a predicate $q\colon Y \rightarrow
[0,1]$ on $Y$ we can form a predicate $c \pull q$ on $X$ via:
\[ \begin{array}{rcl}
\big(c \pull q\big)(x)
& \coloneqq &
\displaystyle\sum_{y\in Y} c(x)(y)\cdot q(y).
\end{array} \]

\noindent There is then a basic equality of validities: $c \push
\omega \models q \;=\; \omega \models c \pull q$.

Given a channel $c \colon X \chanto Y$ and a `prior' state
$\omega\in\Dst(Y)$ we can obtain a reversed channel $c_{\omega}^{\dag}
\colon Y \chanto X$, with $\omega$ as prior distribution, called the
\emph{dagger} of $c$, see~\cite{ClercDDG17,ChoJ19,Fritz20}. Its
formulation is:
\begin{equation}
\label{DaggerEqn}
\begin{array}{rcccl}
c_{\omega}^{\dag}(y)
& \coloneqq &
\omega|_{c \pull \indic{y}}
& = &
\displaystyle\sum_{x\in X} \frac{\omega(x)\cdot c(x)(y)}{(c \push\omega)(y)}
  \,\bigket{x}.
\end{array}
\end{equation}

\noindent This definition uses the point predicate $\indic{y} \colon Y
\rightarrow [0,1]$ which is $1$ on $y\in Y$ and $0$ everywhere
else. The definition only works when the pushforward distribution $c
\push \omega$ has full support. The dagger is also called Bayesian
inversion; it corresponds to turning a conditional probability
$p(y|x)$ into $p(x|y)$.

Daggers of \emph{deterministic} channels play a special role; they
give rise to dagger idempotents.

\begin{lemma}
\label{DetDaggerLem}
Let $f\colon X \rightarrow Y$ be a function, considered as a
(deterministic) channel, and let $\omega\in\Dst(X)$ be a distribution
such that $f \push \omega = \Dst(f)(\omega)$ has full support. Then:
\[ \begin{array}{rclcrclcrcl}
f^{\dag}_{\omega} \push \Dst(f)(\omega)
& = &
\omega
& \qquad\qquad &
f \klafter f^{\dag}_{\omega}
& = &
\idmap
& \qquad\qquad &
\big(f^{\dag}_{\omega} \klafter f\big)^{\dag}_{\omega}
& = &
f^{\dag}_{\omega} \klafter f.
\end{array} \]

\noindent This makes $f$ a \emph{deterministic dagger epi}, with prior
$\omega$, and $f^{\dag}_{\omega} \klafter f$ a \emph{split dagger
  idempotent}, in $\Kl(\Dst)$.
\end{lemma}

\begin{myproof}
We only do the middle equation and leave the other equations to the reader.
\[ \begin{array}{rcccccccl}
\big(f \klafter f^{\dag}_{\omega}\big)(y)
& = &
\displaystyle\sum_{x\in X} f^{\dag}_{\omega}(y)(x)\,\bigket{f(x)}
& \smash{\stackrel{\eqref{DaggerEqn}}{=}} &
\displaystyle\sum_{x\in f^{-1}(y)} \frac{\omega(x)}{\Dst(f)(\omega)(y)}
   \,\bigket{f(x)}
& = &
\displaystyle\frac{\sum_{x\in f^{-1}(y)}\omega(x)}{\Dst(f)(\omega)(y)}\,\bigket{y}
& = &
1\ket{y}.
\end{array} \eqno{\QEDbox} \]

\auxproof{
\[ \begin{array}{rcl}
\big(f^{\dag}_{\omega} \push \Dst(f)(\omega)\big)(x)
& = &
\displaystyle \sum_{y\in Y} \Dst(f)(\omega)(y) \cdot f^{\dag}_{\omega}(y)(x)
\\
& = &
\displaystyle \sum_{y\in Y} \Dst(f)(\omega)(y) \cdot 
   \frac{\omega(x)\cdot \klin{f}(x)(y)}{(\klin{f} \push\omega)(y)}
\\
& = &
\omega(x).
\end{array} \]

Let's abbreviate $e \coloneqq f^{\dag}_{\omega} \after f\colon X \rightarrow X$
for the split idempotent. It is a dagger idempotent, since:
\[ \begin{array}{rcl}
e^{\dag}_{\omega}(x)
& \smash{\stackrel{\eqref{DaggerEqn}}{=}} &
\displaystyle\sum_{y\in X} \frac{\omega(y)\cdot f^{\dag}_{\omega}\big(f(y)\big)(x)}
   {((f^{\dag}_{\omega} \after f) \push\omega)(x)} \,\bigket{y}
\\
& = &
\displaystyle\sum_{y\in X} \frac{\omega(y)\cdot 
   \frac{\omega(x)\cdot \klin{f}(x)(y)}{\Dst(f)(\omega)(f(x))}}
   {(f^{\dag}_{\omega}\push\Dst(f)(\omega)(x)} \,\bigket{y}
\\
& = &
\displaystyle\sum_{y\in f^{-1}(f(x))} \frac{\omega(y)\cdot 
   \frac{\omega(x)}{\Dst(f)(\omega)(f(x))}}
   {\omega(x)} \,\bigket{y}
\\
& = &
\displaystyle\sum_{y\in f^{-1}(f(x))} \frac{\omega(y)}
   {\Dst(f)(\omega)(f(y))} \,\bigket{y}
\\
& = &
f^{\dag}_{\omega}\big(f(y)\big)(x)
\\
& = &
e(x).
\end{array} \]

\noindent Alternatively, using functoriality of dagger:
\[ \begin{array}{rcccl}
\big(f^{\dag}_{\omega} \after f\big)^{\dag}_{\omega}
& = &
f^{\dag}_{\omega} \klafter \big(f^{\dag}_{\omega}\big)^{\dag}_{f \push \omega}
& = &
f^{\dag}_{\omega} \klafter f.
\end{array} \]
}
\end{myproof}

\subsection{Multiset partitions}\label{PartitionSubsec}

A \emph{multiset partition}, or simply a
\emph{partition}\footnote{Commonly, the word `partition' is used for
`cover': a collection of pairwise disjoint subsets whose union is the
whole set. This may be confusing. We use the phrases `multiset
partition' and `set partition' (cover) to distinguish them
in~\cite{Jacobs22g,JacobsS23a}. Here, only multiset partitions are
used, so we often simply use `partition' for `multiset partition'.},
is a multiset over the non-negative natural numbers $\pNNO$. Such a
partition $\sum_{i}n_{i}\ket{i} \in \Mlt(\pNNO)$ has a \emph{sum} (or
mean, or average) defined as:
\[ \begin{array}{rclcrcl}
\som\Big(\sum_{i}\,n_{i}\ket{i}\Big)
& \coloneqq &
\sum_{i}\, n_{i}\cdot i \;\in\; \pNNO
& \quad\mbox{and used in:}\quad &
\MP(K)
& \coloneqq &
\setin{\varphi}{\Mlt(\pNNO)}{\som(\varphi) = K}.
\end{array} \]

\noindent Thus, $\MP(K)$ is the set of partitions with sum
$K$. The sizes $\big|\,\MP(K)\big|$ of the sets of partitions with
sum $K=1,2,3,\ldots$ are given by the so-called partition
numbers~\cite{Andrews98}: $1, 2, 3, 5, 7, 11, 15, 22, 30, 42, \ldots$
for which no closed-form expression is known. For instance, $\MP(4)$
contains the following five multisets.
\begin{equation}
\label{PartitionsFour}
4\ket{1} \quad 2\ket{1} + 1\ket{2} \quad 1\ket{1} + 1\ket{3}
   \quad 2\ket{2} \quad 1\ket{4}.
\end{equation}

\noindent These partitions can be seen as the different ways to
describe the sum $4$ via coins: 4 coins with value 1, 2 coins of 1 and
1 of 2, \textit{etc}, see~\cite{Jacobs22c}.  Such partitions are
studied for instance in population biology, see
\textit{e.g.}~\cite{Crane16,Ewens72,Joyce98,Kingman78a,Kingman78b} and
also Section~\ref{SizeSuffStatSec}. In economics they can be used for
(un)fairness~\cite{Kolm69}: if you have $4$ units of `wealth' that you
can distribute over $4$ people, you can given each of them one unit
(as on left in~\eqref{PartitionsFour}) or you can give all units to
one individual (as on the right in~\eqref{PartitionsFour}). The
partitions in the middle are intermediate forms of fairness. Notice
that the number $\som(\varphi)$, for $\varphi\in\Mlt(\pNNO)$, is
typically different from its size $\|\varphi\|$. We do have
$\|\varphi\| \leq \som(\varphi)$. For instance, the partitions
in~\eqref{PartitionsFour} all have sum four, but their sizes are, in
order, $4, 3, 2, 2, 1$.

The next result involves a \emph{multiplicity count} function
$\mulcount$, introduced in~\cite{Jacobs22c} (and used also
in~\cite{Jacobs22g,JacobsS23a}), that plays a fundamental role in the
sequel --- as analogue of accumulation $\acc$. It counts the
multiplicities in a multiset.  For instance,
\[ \begin{array}{rcl}
\mulcount\Big(3\ket{a} + 1\ket{b} + 3\ket{c} + 2\ket{d}\Big)
& = &
1\ket{1} + 1\ket{2} + 2\ket{3}.
\end{array} \]

\noindent This expresses that in the above (argument) multiset over
$\{a,b,c,d\}$, 1 element occurs 1 time (namely $b$), 1 element occurs
2 times (namely $d$) and 2 elements occur 3 times (namely $a$ and
$c$). Thus, the multiplicity count function abstracts away from the
elements in a multiset and only looks at how many elements occur how
many times. It is a dagger epi, see Section~\ref{MulcountSuffStatSec}.

The next result collects some combinatorial results about the relation
between multisets and partitions, via this new multiplicity count
function $\mulcount$. The focus is on describing $\mulcount$ as the
retraction part of a split idempotent.

\begin{proposition}
\label{MulcountProp}
Let $X$ be a finite set with at least $n \geq K$ elements, that is,
with $n \coloneqq \setsize{X} \geq K$.
\begin{enumerate}
\item \label{MulcountPropMulcount} Define for a multiset
  $\varphi\in\Mlt(X)$, its multiplicity count as: $\mulcount(\varphi)
  \coloneqq \displaystyle\sum_{x\in\supp(\varphi)}
  1\bigket{\varphi(x)}$.  This function restricts to $\mulcount \colon
  \Mlt[K](X) \rightarrow \MP(K)$.

\item \label{MulcountPropSize} For each $\sigma\in\MP(K)$ there are
\[ \begin{array}{rcl}
\displaystyle\binom{n}{\sigma}
& = &
\displaystyle\frac{n!}{\facto{\sigma}\cdot 
   (n-\|\sigma\|)!} \quad
  \mbox{many multisets $\varphi\in\Mlt[K](X)$ with 
   $\mulcount(\varphi) = \sigma$.}
\end{array} \]

\noindent This means: $\bigsetsize{\Mlt[K](X) \cap
  \mulcount^{-1}(\sigma)} = \displaystyle\binom{n}{\sigma}$.

\item \label{MulcountPropStack} Thus, we can define a \emph{stack}
  channel $\stack \colon \MP(K) \chanto \Mlt[K](X)$ via uniform
  distributions as:
\begin{equation}
\label{StackEqn}
\begin{array}{rcl}
\stack(\sigma)
& \coloneqq &
\displaystyle\sum_{\varphi\in\mulcount^{-1}(\sigma)} \, 
   \frac{1}{\binom{n}{\sigma}} \,\bigket{\varphi}.
\end{array}
\end{equation}

\noindent Then: $\mulcount \after \stack = \idmap \colon \MP(K)
\chanto \MP(K)$.

\item \label{MulcountPropPermute} Write $\eltperm \coloneqq \stack
  \klafter \mulcount \colon \Mlt[K](X) \chanto \Mlt[K](X)$ for the
  resulting split idempotent in $\Kl(\Dst)$, where $\eltperm$ stands
  for element permutation (``substitution''). It satisfies, for
  $\varphi\in\Mlt[K](X)$,
\[ \begin{array}{rcccl}
\eltperm(\varphi)
& = &
\displaystyle\sum_{\psi\in\mulcount^{-1}(\mulcount(\varphi))} \, 
   \frac{1}{\binom{n}{\mulcount(\varphi)}} \,\bigket{\psi}
& = &
\displaystyle\sum_{\pi\in\Perm(X)} \frac{1}{n!}
   \, \Bigket{\Mlt(\pi)(\varphi)}.
\end{array} \]

\noindent The first equation holds by definition; the second one is
the statement.
\end{enumerate}
\end{proposition}

\begin{myproof}
\begin{enumerate}
\item Straightforward.

\item By assumption, the set $X$ has $n$ elements. The question is:
  how many ways are there to partition this set $X$ into subsets with
  $\sigma(1)$, $\sigma(2)$, \ldots, $\sigma(K)$ elements, and finally,
  with $n-\|\sigma\|$ many remaining elements? As we have seen in
  Lemma~\ref{AccLem}~\eqref{AccLemSize} this number of partitions is
  given by the multinomial coefficient:
\[ \begin{array}{rcccccl}
\displaystyle\binom{n}{\sigma(1), \ldots, \sigma(K), n\!-\!\|\sigma\|}
& = &
\displaystyle\frac{n!}{\sigma(1)! \cdot \ldots \cdot \sigma(K)! \cdot
   (n\!-\!\|\sigma\|)!}
& = &
\displaystyle\frac{n!}{\facto{\sigma}  \cdot (n\!-\!\|\sigma\|)!}
& = &
\displaystyle\binom{n}{\sigma}.
\end{array} \]

\item For a partition $\sigma\in\MP(K)$ with sum $K$,
\[ \begin{array}{rcccccl}
\big(\mulcount \klafter \stack\big)(\sigma)
& \smash{\stackrel{\eqref{StackEqn}}{=}} &
\displaystyle\sum_{\varphi\in\mulcount^{-1}(\sigma)} \, 
   \frac{1}{\binom{n}{\sigma}} \,\bigket{\mulcount(\varphi)}
& = &
\displaystyle \left(\sum_{\varphi\in\mulcount^{-1}(\sigma)} \, 
   \frac{1}{\binom{n}{\sigma}}\right)\bigket{\sigma}
& = &
1\bigket{\sigma}.
\end{array} \]

\noindent This last equation follows from item~\eqref{MulcountPropSize}.

\item For $\varphi\in\Mlt[K](X)$ write $\sigma =
  \mulcount(\varphi)$. It suffices to prove:
\[ \begin{array}{rcl}
\displaystyle\sum_{\psi\in\mulcount^{-1}(\sigma)} \facto{\sigma}\cdot
   \big(n-\|\sigma\|\big)! \, \bigket{\psi}
& = &
\displaystyle\sum_{\pi\in\Perm(X)} 1\Bigket{\Mlt(\pi)(\varphi)}.
\end{array} \eqno{(*)} \]

\noindent For a permutation $\pi \colon X \congrightarrow X$ the
multiset $\Mlt(\pi)(\varphi)$ obtained by permuting the elements
occurring in $\varphi$, according to $\pi$, also has multiplicity
count $\sigma = \mulcount(\varphi)$. Some of these permutations of
$\varphi$ are equal to $\varphi$. We have to count their number.

For each $i\in\NNO$ with $\sigma(i) > 0$ there are $\sigma(i)$ many
elements of $X$ in $\varphi$ occurring $i$ times. Permuting these
$\sigma(i)$ many elements does not change the multiset $\varphi$.
This can be done in $\sigma(i)!$ many ways. Using this argument for
each $i$ explains the occurrence of $\facto{\sigma} =
\prod_{i}\sigma(i)!$ in~$(*)$. There are $n-\|\sigma\| =
\big|\supp(\varphi)\big|$ many elements of $X$ that do not occur in
$\varphi$. They can be permuted in $(n-\|\sigma\|)!$ many ways without
changing $\varphi$. \QED
\end{enumerate}
\end{myproof}

This last item~\eqref{MulcountPropPermute} is somewhat mysterious, so
we like to illustrate what is going on.

\begin{example}
\label{MulCountEx}
Consider a set $X = \{a,b,c,d\}$ with a multiset $\varphi = 2\ket{a} +
1\ket{b} + 1\ket{c} + 1\ket{d}$ of size $5$, whose multiplicity count
is $\sigma \coloneqq \mulcount(\varphi) = 3\ket{1} + 1\ket{2}$, with
sum $5$. The multisets $\psi\in\Mlt[5](X)$ with $\mulcount(\psi) =
\sigma$ are:
\[ \begin{array}{ccc}
2\ket{a} + 1\ket{b} + 1\ket{c} + 1\ket{d}, 
& \hspace*{3em} &
1\ket{a} + 2\ket{b} + 1\ket{c} + 1\ket{d}, 
\\
1\ket{a} + 1\ket{b} + 2\ket{c} + 1\ket{d}, 
& &
1\ket{a} + 1\ket{b} + 1\ket{c} + 2\ket{d}. 
\end{array} \]

\noindent There are $4! = 24$ permutations $\pi$ of $X$. The
`swapped' multiset $\Mlt(\pi)(\varphi)$ is of one of these four forms.
Each of these forms occurs $6$ times in:
\[ \begin{array}{rcl}
\displaystyle\sum_{\pi\in\sigma(X)} \frac{1}{24}\Bigket{\Mlt(\pi)(\varphi)}
& = &
\displaystyle\frac{1}{4}\Bigket{2\ket{a} + 1\ket{b} + 1\ket{c} + 1\ket{d}} + 
\frac{1}{4}\Bigket{1\ket{a} + 2\ket{b} + 1\ket{c} + 1\ket{d}} \, +
\\[-0.3em]
& & \quad
\displaystyle\frac{1}{4}\Bigket{1\ket{a} + 1\ket{b} + 2\ket{c} + 1\ket{d}} +
\frac{1}{4}\Bigket{1\ket{a} + 1\ket{b} + 1\ket{c} + 2\ket{d}}
\hspace*{\arraycolsep}=\hspace*{\arraycolsep}
\eltperm\big(\varphi\big).
\end{array} \]

\noindent This probability $\frac{1}{4}$ equals
$\frac{\sfacto{\sigma}\cdot (|X|-\|\sigma\|)!}{24} = \frac{3!\cdot 1!
  \cdot 0!}{24}$, as prescribed in
Proposition~\ref{MulcountProp}~\eqref{MulcountPropPermute}.
\end{example}

%% 2|a> + 1|b> + 1|c> + 1|d>
%% 3|1> + 1|2>

%% 6|2|a> + 1|b> + 1|c> + 1|d>> + 
%% 6|1|a> + 2|b> + 1|c> + 1|d>> + 
%% 6|1|a> + 1|b> + 2|c> + 1|d>> + 
%% 6|1|a> + 1|b> + 1|c> + 2|d>>

%% 6

%% 1|2|a> + 1|b> + 1|c> + 1|d>> + 
%% 1|1|a> + 2|b> + 1|c> + 1|d>> + 
%% 1|1|a> + 1|b> + 2|c> + 1|d>> + 
%% 1|1|a> + 1|b> + 1|c> + 2|d>>

\section{Sufficient statistic}\label{SuffStatSec}

The diagrammatic formulation of the notion of sufficient statistic has
been introduced recently in~\cite{Fritz20}. This formulation involves
a simple equation between two diagrams, see~\eqref{SuffStatEqn} below.
Implicitly, it relies on disintegration, which we introduce first
(based on~\cite{Fritz20} and also~\cite{ChoJ19}).  It helps to
understand sufficiency as a form of updating that trivialises, by
making a parameter obsolete.

\begin{definition}
\label{DisintDef}
Let $c \colon A \chanto X\times Y$ be a channel, represented as a box
on the left below. A \emph{disintegration} of $c$ is a channel
$d\colon A\times Y \chanto X$ that satisfies the equation on the
right.
\begin{equation}
\label{DisintEqn}
\vcenter{\hbox{\begin{tikzpicture}
	\begin{pgfonlayer}{nodelayer}
		\node [style=medium box] (0) at (0, 0) {$c$};
		\node [style=none] (1) at (-0.5, 0.5) {};
		\node [style=none] (2) at (0.5, 0.5) {};
		\node [style=none] (3) at (0.5, 1.5) {};
		\node [style=none] (4) at (-0.5, 1.5) {};
		\node [style=none] (5) at (-1, 1.5) {$X$};
		\node [style=none] (6) at (1, 1.5) {$Y$};
		\node [style=none] (7) at (0.5, -1.25) {$A$};
		\node [style=none] (8) at (0, -1.25) {};
	\end{pgfonlayer}
	\begin{pgfonlayer}{edgelayer}
		\draw (3.center) to (2.center);
		\draw (4.center) to (1.center);
		\draw (8.center) to (0);
	\end{pgfonlayer}
\end{tikzpicture}}}
\hspace*{10em}
\vcenter{\hbox{\begin{tikzpicture}
	\begin{pgfonlayer}{nodelayer}
		\node [style=medium box] (0) at (-4.15, 0) {$c$};
		\node [style=none] (1) at (-4.65, 0.5) {};
		\node [style=none] (2) at (-3.65, 0.5) {};
		\node [style=none] (3) at (-3.65, 3) {};
		\node [style=none] (4) at (-4.65, 3) {};
		\node [style=none] (8) at (-4.15, -2.75) {};
		\node [style=medium box] (9) at (1, -0.75) {$c$};
		\node [style=none] (10) at (0.5, -0.25) {};
		\node [style=none] (11) at (1.5, -0.25) {};
		\node [style=none] (13) at (0.75, 1.5) {};
		\node [style=black dot] (14) at (1, -2) {};
		\node [style=none] (15) at (1, -2.75) {};
		\node [style=medium box] (16) at (0.25, 2) {$d$};
		\node [style=none] (17) at (-0.25, 1.5) {};
		\node [style=none] (18) at (-0.25, -1) {};
		\node [style=upground] (19) at (0.5, 0.5) {};
		\node [style=none] (21) at (0.25, 3) {};
		\node [style=black dot] (22) at (1.5, 0.5) {};
		\node [style=none] (23) at (1.5, 3) {};
		\node [style=none] (24) at (-2, 0) {$=$};
	\end{pgfonlayer}
	\begin{pgfonlayer}{edgelayer}
		\draw (3.center) to (2.center);
		\draw (4.center) to (1.center);
		\draw (8.center) to (0);
		\draw (10.center) to (19);
		\draw (14) to (9);
		\draw (15.center) to (14);
		\draw (18.center) to (17.center);
		\draw [in=150, out=-90, looseness=1.25] (18.center) to (14);
		\draw (21.center) to (16);
		\draw [in=150, out=-90, looseness=1.25] (13.center) to (22);
		\draw (22) to (23.center);
		\draw (11.center) to (22);
	\end{pgfonlayer}
\end{tikzpicture}}}
\end{equation}

\noindent In this diagram we use $\ground$ for discarding
(marginalisation, projecting away) and $\minicopy$ for copying. On the
right-hand-side of the equation, the $X$-output of the $c$ channel is
discarded, but reconstructed via the $d$-channel.
\end{definition}

%% We can alternatively formulate Equation~\eqref{DisintEqn} as:
%% \[ \begin{array}{rcl}
%% c
%% & = &
%% (d\otimes\idmap) \klafter (\idmap\otimes\Delta) \klafter 
%%    (\idmap\otimes\pi_{2}) \klafter (\idmap\otimes c) \klafter \Delta.
%% \end{array} \]

%% \noindent The above string diagrammatic formulation better captures
%% what is going. 

We do not worry about uniqueness of disintegration in the current
setting. It does exist, under additional requirements,
see~\cite{Fritz20,ChoJ19} for details.

We use surjective functions $s\colon X \twoheadrightarrow Y$ to
identify elements in $X$, resulting in abstractions in $Y$. This $Y$
may give numerical information, \textit{e.g.}~when $Y$ is the set of
natural or real numbers. Such a map is also called a
\emph{statistic}. A statistic $s$ is called \emph{sufficient} for a
channel, as statistical model depending on a parameter, when the
dependence on the parameter disappears via updating through
$s$. Sufficency refers to being adequate as a summary of essential
aspects of the elements in $X$. Sufficiency involves the existence of
a reversal of the \emph{function} $s\colon X \rightarrow Y$ to a
\emph{channel} $d\colon Y \chanto X$. This channel $d$ can be used to
reconstruct a distribution $X$, form its summary given by $s$. We
first give a string diagrammatic description,
following~\cite[Defn.~14.3]{Fritz20}. Subsequently we explain the
relation to disintegration.

\begin{definition}
\label{SuffStatDef}
Let $c\colon A \rightarrow X$ be a channel, where we think
of $A$ as the space of parameters.
\begin{enumerate}
\item A \emph{statistic} for the channel $c$ is a function $s\colon X
  \rightarrow Y$.

\item Such a statistic $s$ is \emph{sufficient} if there is a channel
  $d\colon Y \chanto X$ such that:
\begin{equation}
\label{SuffStatEqn}
\vcenter{\hbox{\begin{tikzpicture}
	\begin{pgfonlayer}{nodelayer}
		\node [style=black dot] (25) at (4.75, -1) {};
		\node [style=none] (26) at (3.75, 2.5) {};
		\node [style=small box] (27) at (4.75, -2) {$c$};
		\node [style=small box] (28) at (5.75, 0.5) {$s$};
		\node [style=none] (29) at (3.75, 0.5) {};
		\node [style=none] (30) at (4.75, -3.25) {};
		\node [style=none] (31) at (5.75, 2.5) {};
		\node [style=black dot] (32) at (10.5, 0.25) {};
		\node [style=none] (33) at (11.5, 2.5) {};
		\node [style=small box] (34) at (10.5, -2.25) {$c$};
		\node [style=small box] (35) at (9.5, 1.5) {$d$};
		\node [style=none] (36) at (11.5, 1.5) {};
		\node [style=none] (37) at (10.5, -3.25) {};
		\node [style=none] (38) at (9.5, 2.5) {};
		\node [style=none] (39) at (7.75, -0.5) {$=$};
		\node [style=small box] (40) at (10.5, -0.75) {$s$};
	\end{pgfonlayer}
	\begin{pgfonlayer}{edgelayer}
		\draw (25) to (27);
		\draw [in=-90, out=30, looseness=1.25] (25) to (28);
		\draw (26.center) to (29.center);
		\draw [in=150, out=-90, looseness=1.25] (29.center) to (25);
		\draw (30.center) to (27);
		\draw (28) to (31.center);
		\draw [in=-90, out=165] (32) to (35);
		\draw (33.center) to (36.center);
		\draw (37.center) to (34);
		\draw (35) to (38.center);
		\draw [in=-90, out=15] (32) to (36.center);
		\draw (34) to (40);
		\draw (40) to (32);
	\end{pgfonlayer}
\end{tikzpicture}}}
\end{equation}
\end{enumerate}
\end{definition}

\begin{remark}
\label{SuffStatRem}
\begin{enumerate}
\item The channel $d$ on the right-hand-side in~\eqref{SuffStatEqn}
  results in a special way from disintegration, namely disintegration
  of the channel $\tuple{\idmap,s} \klafter c$ on the left-hand-side
  in~\eqref{SuffStatEqn}. If we write $D \colon A\times Y \chanto X$
  for its disintegration channel, as in~\eqref{DisintEqn}, then, we
  should get an equation as given on the left below. The
  distinguishing property of a sufficient statistic is that the dashed
  line is absent.
\begin{equation}
\label{SuffStatDisintEqn}
\vcenter{\hbox{\begin{tikzpicture}
	\begin{pgfonlayer}{nodelayer}
		\node [style=black dot] (0) at (-3, -1) {};
		\node [style=none] (1) at (-4, 2.5) {};
		\node [style=small box] (2) at (-3, -2) {$c$};
		\node [style=small box] (3) at (-2, 0.5) {$s$};
		\node [style=none] (4) at (-4, 0.5) {};
		\node [style=none] (5) at (-3, -3.25) {};
		\node [style=none] (6) at (-2, 2.5) {};
		\node [style=black dot] (7) at (3.5, 0.75) {};
		\node [style=none] (8) at (4.5, 3) {};
		\node [style=small box] (9) at (3.5, -1.75) {$c$};
		\node [style=none] (11) at (4.5, 2) {};
		\node [style=none] (13) at (2.25, 3) {};
		\node [style=none] (14) at (0, -0.5) {$=$};
		\node [style=small box] (15) at (3.5, -0.25) {$s$};
		\node [style=medium box] (16) at (2.25, 2) {$D$};
		\node [style=none] (17) at (1.75, 1.75) {};
		\node [style=none] (18) at (2.75, 1.75) {};
		\node [style=black dot] (19) at (3.5, -3.25) {};
		\node [style=none] (20) at (1.75, -1.5) {};
		\node [style=none] (22) at (3.5, -4) {};
	\end{pgfonlayer}
	\begin{pgfonlayer}{edgelayer}
		\draw (0) to (2);
		\draw [in=-90, out=30, looseness=1.25] (0) to (3);
		\draw (1.center) to (4.center);
		\draw [in=150, out=-90, looseness=1.25] (4.center) to (0);
		\draw (5.center) to (2);
		\draw (3) to (6.center);
		\draw (8.center) to (11.center);
		\draw [in=-90, out=15] (7) to (11.center);
		\draw (9) to (15);
		\draw (15) to (7);
		\draw (16) to (13.center);
		\draw [in=150, out=-90, looseness=1.25] (18.center) to (7);
		\draw (19) to (9);
		\draw (22.center) to (19);
		\draw [style=dashed edge, in=-90, out=150] (19) to (20.center);
		\draw [style=dashed edge] (20.center) to (17.center);
	\end{pgfonlayer}
\end{tikzpicture}}}
\hspace*{10em}
\vcenter{\hbox{\begin{tikzpicture}
	\begin{pgfonlayer}{nodelayer}
		\node [style=none] (0) at (-2.3, 2.075) {};
		\node [style=medium box] (1) at (-2.3, 0.825) {$D$};
		\node [style=none] (2) at (-2.8, 0.575) {};
		\node [style=none] (3) at (-1.8, 0.575) {};
		\node [style=none] (4) at (-2.75, -0.5) {};
		\node [style=none] (5) at (-1.75, -0.5) {};
		\node [style=none] (6) at (1.75, -0.5) {};
		\node [style=none] (7) at (3.25, -0.5) {};
		\node [style=upground] (8) at (1.75, 0.75) {};
		\node [style=none] (9) at (3.25, 2) {};
		\node [style=none] (10) at (0, 0.75) {$=$};
		\node [style=small box] (11) at (3.25, 0.75) {$d$};
	\end{pgfonlayer}
	\begin{pgfonlayer}{edgelayer}
		\draw (1) to (0.center);
		\draw (2.center) to (4.center);
		\draw (3.center) to (5.center);
		\draw (8) to (6.center);
		\draw (11) to (9.center);
		\draw (11) to (7.center);
	\end{pgfonlayer}
\end{tikzpicture}}}
\end{equation}

\noindent The situation on the left can be expressed by the equation
on the right.

\item The fact that the dashed wire is missing
  in~\eqref{SuffStatDisintEqn} is often expressed as: the conditional
  distribution $c(a)$, given $s$, does not depend on $a$. This can be
  made more precise. The box/channel $D$ in~\eqref{SuffStatDisintEqn}
  can be calculated --- in discrete probability --- as a distribution
  $D(a,y)\in\Dst(X)$, for $a\in A$ and $y\in Y$, via a dagger (see
  Subsection~\ref{PredicateSubsec}).  Explicitly,
\begin{equation}
\label{SuffStatDaggerEqn}
\begin{array}{rcccccl}
D(a,y)
& = &
s_{c(a)}^{\dag}(y)
& \smash{\stackrel{\eqref{DaggerEqn}}{=}} &
c(a)|_{s \pull \indic{y}}
& = &
\displaystyle\sum_{x\in s^{-1}(y)} \frac{c(a)(x)}{\Dst(s)(c(a))(y)}\,\bigket{x}.
\end{array}
\end{equation}

\noindent The absence of the dashed arrow in~\eqref{SuffStatDisintEqn}
corresponds to the non-dependence of the latter expressions on $a$.
This is the essence of the Fisher-Neyman factorisation theorem,
see~\cite[Thm.~14.5]{Fritz20} and~\cite[Prop~4.10]{BernardoS00}
or~\cite[\S3.3]{SuhovK05}.

\item In the sequel we are interested in actually demonstrating
  sufficiency of certain statistics, in discrete probability. Proving
  Equation~\ref{SuffStatEqn} amounts to showing an equality of
  distributions of the following form. For each parameter $a\in A$,
\begin{equation}
\label{SuffStatKetEqn}
\begin{array}{rcl}
\tuple{\idmap,s} \push c(a)
& = &
\displaystyle \sum_{x\in X} c(a)(x)\,\Bigket{x, s(x)}
\\
& = &
\displaystyle\sum_{x,z\in X} c(a)(z)\cdot d\big(s(z)\big)(z) \, \Bigket{z,s(x)}
\hspace*{\arraycolsep}=\hspace*{\arraycolsep}
\tuple{d,\idmap} \push \Dst(s)\big(c(a)\big).
\end{array}
\end{equation}

\auxproof{
\noindent This follows since the above left-hand-side equals the
left-hand-side in~\ref{SuffStatEqn} and the above right-hand-side
equals the right-hand-side in~\ref{SuffStatEqn} since:
\[ \begin{array}{rcl}
\displaystyle \sum_{x\in X, y\in Y} \Dst(s)\big(c(a))(y) \cdot d(y)(x) \, 
   \Bigket{x,y}
& = &
\displaystyle \sum_{x\in X, y\in Y} \left(\sum_{z\in s^{-1}(y)} c(a)(z)\right) 
   \cdot d(y)(x) \, \Bigket{x,y}
\\[+0.5em]
& = &
\displaystyle\sum_{x\in X, y\in Y, z\in s^{-1}(y)} c(a)(z)\cdot d(y)(z) \, \Bigket{x,y}.
\\[+0.5em]
& = &
\displaystyle\sum_{x,z\in X} c(a)(z)\cdot d\big(s(z)\big)(z) \, \Bigket{z,s(x)}.
\end{array} \]
}

\item Two distributions $\omega,\rho\in\Dst(X)$ are equal when
  $\omega\models p \;=\; \rho\models p$ for all predicates $p$ on $X$.
  This easily follows by just looking at point predicates $\indic{x}$,
  where $x\in X$, for which $\omega\models\indic{x} = \omega(x)$.

This validity formulation of equality of distributions can also be
used for the sufficiency equation~\eqref{SuffStatEqn}. It amounts to,
for all predicates $p\in\Pred(X)$, $q\in\Pred(Y)$,
\begin{equation}
\label{SuffStatPredEqn}
\begin{array}{rcl}
c(a) \models p \andthen (s \pull q)
& \;=\; &
\Dst(s)\big(c(a)\big) \models (d \pull p) \andthen q.
\end{array}
\end{equation}

\noindent This formulation uses conjunction $\andthen$ for pointwise
multiplication of predicates $(p_{1}\andthen p_{2})(x) = p_{1}(x)
\cdot p_{2}(x)$. It has a clear adjointness flavour. However, there is
no direction involved, or left-right distinction, since $\andthen$ is
a commutative operation --- in classical probability, not in quantum
probability, see
\textit{e.g.}~\cite{ChoJWW15b,Jacobs15d,WesterbaanB19}.
\end{enumerate}
\end{remark}

All our examples below follow the following pattern to obtain
a sufficient statistic.

\begin{lemma}
\label{SplitIdempotentSuffStatLem}
Consider a channel together with a split idempotent in $\Kl(\Dst)$,
consisting of a section followed by a retraction. Then:
\begin{equation}
\label{SplitIdempotentSuffStatEqn}
\vcenter{\hbox{\begin{tikzpicture}
	\begin{pgfonlayer}{nodelayer}
		\node [style=black dot] (25) at (10.25, 0.5) {};
		\node [style=none] (26) at (9, 4) {};
		\node [style=small box] (27) at (10.25, -0.5) {$\text{chan}$};
		\node [style=small box] (28) at (11.5, 2) {$\text{retraction}$};
		\node [style=none] (29) at (9, 2) {};
		\node [style=none] (30) at (10.25, -1.75) {};
		\node [style=none] (31) at (11.5, 4) {};
		\node [style=black dot] (32) at (17.75, 1.75) {};
		\node [style=none] (33) at (19, 4.25) {};
		\node [style=small box] (34) at (17.75, -0.75) {$\text{chan}$};
		\node [style=small box] (35) at (16.5, 3) {$\text{section}$};
		\node [style=none] (36) at (19, 3) {};
		\node [style=none] (37) at (17.75, -2) {};
		\node [style=none] (38) at (16.5, 4.25) {};
		\node [style=none] (39) at (14.25, 1.25) {$=$};
		\node [style=small box] (40) at (17.75, 0.75) {$\text{retraction}$};
		\node [style=medium box] (41) at (-4, 2) {\raisebox{0.1em}{$\begin{array}{c} \text{split} \\[-0.8em] \text{idempotent} \end{array}$}};
		\node [style=medium box] (42) at (-4, 0.25) {$\text{chan}$};
		\node [style=none] (43) at (-4, 3.5) {};
		\node [style=none] (44) at (-4, -1.25) {};
		\node [style=none] (45) at (-1, 1.25) {$=$};
		\node [style=medium box] (46) at (1.25, 1.25) {$\text{chan}$};
		\node [style=none] (47) at (1.25, -1.25) {};
		\node [style=none] (48) at (1.25, 3.5) {};
		\node [style=none] (49) at (5.5, 1.25) {$\text{implies}$};
	\end{pgfonlayer}
	\begin{pgfonlayer}{edgelayer}
		\draw (25) to (27);
		\draw [in=-90, out=30, looseness=1.25] (25) to (28);
		\draw (26.center) to (29.center);
		\draw [in=150, out=-90, looseness=1.25] (29.center) to (25);
		\draw (30.center) to (27);
		\draw (28) to (31.center);
		\draw [in=-90, out=165] (32) to (35);
		\draw (33.center) to (36.center);
		\draw (37.center) to (34);
		\draw (35) to (38.center);
		\draw [in=-90, out=15] (32) to (36.center);
		\draw (34) to (40);
		\draw (40) to (32);
		\draw (42) to (44.center);
		\draw (42) to (41);
		\draw (41) to (43.center);
		\draw (46) to (47.center);
		\draw (46) to (48.center);
	\end{pgfonlayer}
\end{tikzpicture}}}
\end{equation}

\noindent The latter equation says that the retraction is a sufficient
statistic for the channel.
\end{lemma}

\begin{myproof}
This is in essence a special case of the Fisher-Neyman factorisation
theorem, as formulated in diagrammatic form
in~\cite[Thm.~14.5]{Fritz20}. It uses that $\Kl(\Dst)$ is a positive
Markov category, justifying the last equation below,
see~\cite[Rem.11.23]{Fritz20}.
\[ \vcenter{\hbox{\begin{tikzpicture}
	\begin{pgfonlayer}{nodelayer}
		\node [style=black dot] (0) at (-7.25, -0.25) {};
		\node [style=none] (1) at (-8.5, 3) {};
		\node [style=small box] (2) at (-7.25, -1.25) {$\text{chan}$};
		\node [style=small box] (3) at (-6, 1.25) {$\text{retraction}$};
		\node [style=none] (4) at (-8.5, 1.25) {};
		\node [style=none] (5) at (-7.25, -2.25) {};
		\node [style=none] (6) at (-6, 3) {};
		\node [style=black dot] (7) at (0.5, 0.75) {};
		\node [style=none] (8) at (-0.75, 3.5) {};
		\node [style=small box] (9) at (0.5, -1.75) {$\text{chan}$};
		\node [style=small box] (10) at (1.75, 2.25) {$\text{retraction}$};
		\node [style=none] (11) at (-0.75, 2.25) {};
		\node [style=none] (12) at (0.5, -2.75) {};
		\node [style=none] (13) at (1.75, 3.5) {};
		\node [style=medium box] (14) at (0.5, -0.25) {\raisebox{0.1em}{$\begin{array}{c} \text{split} \\[-0.8em] \text{idempotent} \end{array}$}};
		\node [style=none] (15) at (-3.25, 0) {$=$};
		\node [style=none] (16) at (4.5, 0) {$=$};
		\node [style=black dot] (17) at (8, 0.75) {};
		\node [style=none] (18) at (6.75, 3.5) {};
		\node [style=small box] (19) at (8, -3.25) {$\text{chan}$};
		\node [style=small box] (20) at (9.25, 2.25) {$\text{retraction}$};
		\node [style=none] (21) at (6.75, 2.25) {};
		\node [style=none] (22) at (8, -4.25) {};
		\node [style=none] (23) at (9.25, 3.5) {};
		\node [style=small box] (24) at (8, -0.25) {$\text{section}$};
		\node [style=small box] (25) at (8, -1.75) {$\text{retraction}$};
		\node [style=black dot] (26) at (15.5, 0.5) {};
		\node [style=none] (27) at (16.75, 3) {};
		\node [style=small box] (28) at (15.5, -2) {$\text{chan}$};
		\node [style=small box] (29) at (14.25, 1.75) {$\text{section}$};
		\node [style=none] (30) at (16.75, 1.75) {};
		\node [style=none] (31) at (15.5, -3) {};
		\node [style=none] (32) at (14.25, 3) {};
		\node [style=none] (33) at (12, 0) {$=$};
		\node [style=small box] (34) at (15.5, -0.5) {$\text{retraction}$};
	\end{pgfonlayer}
	\begin{pgfonlayer}{edgelayer}
		\draw (0) to (2);
		\draw [in=-90, out=30, looseness=1.25] (0) to (3);
		\draw (1.center) to (4.center);
		\draw [in=150, out=-90, looseness=1.25] (4.center) to (0);
		\draw (5.center) to (2);
		\draw (3) to (6.center);
		\draw [in=-90, out=30, looseness=1.25] (7) to (10);
		\draw (8.center) to (11.center);
		\draw [in=150, out=-90, looseness=1.25] (11.center) to (7);
		\draw (12.center) to (9);
		\draw (10) to (13.center);
		\draw (7) to (14);
		\draw (14) to (9);
		\draw [in=-90, out=30, looseness=1.25] (17) to (20);
		\draw (18.center) to (21.center);
		\draw [in=150, out=-90, looseness=1.25] (21.center) to (17);
		\draw (22.center) to (19);
		\draw (20) to (23.center);
		\draw (17) to (24);
		\draw (24) to (25);
		\draw (25) to (19);
		\draw [in=-90, out=165] (26) to (29);
		\draw (27.center) to (30.center);
		\draw (31.center) to (28);
		\draw (29) to (32.center);
		\draw [in=-90, out=15] (26) to (30.center);
		\draw (28) to (34);
		\draw (34) to (26);
	\end{pgfonlayer}
\end{tikzpicture}}} \eqno{\QEDbox} \]
\end{myproof}

This result will be exploited in the next four sections. 

%% There is a trivial way to do so: given an idempotent $e\colon Y
%% \chanto Y$, then for any channel $c\colon X \chanto Y$, the retraction
%% part of $e$ is a sufficient statistic for $e\klafter c$.

\section{Accumulation as sufficient statistic}\label{AccSuffStatSec}

We can now start harvesting from the previous two sections. 

\begin{theorem}
\label{AccSuffStatThm}
Fix a set $X$ and a number $K\in\NNO$.  The accumulation function
$\acc \colon X^{K} \rightarrow \Mlt[K](X)$ is a sufficient statistic
for the identically and independent distribution channel $\iid[K]
\colon \Dst(X) \chanto X^{K}$, via arrangement. This is expressed by
the following equation between channels $\Dst(X) \chanto X^{K} \times
\Mlt[K](X)$.
\begin{equation}
\label{AccSuffStatEqn}
\vcenter{\hbox{\begin{tikzpicture}
	\begin{pgfonlayer}{nodelayer}
		\node [style=medium box] (73) at (2.5, -2) {$\iid[K]$};
		\node [style=none] (74) at (2.5, -3.25) {};
		\node [style=black dot] (75) at (2.5, -0.75) {};
		\node [style=medium box] (77) at (4, 0.75) {$\acc$};
		\node [style=none] (81) at (4, 1.75) {};
		\node [style=none] (82) at (1, 1.75) {};
		\node [style=none] (83) at (1, 0.75) {};
		\node [style=none] (84) at (12.75, 1.75) {};
		\node [style=none] (85) at (9.75, 1.75) {};
		\node [style=medium box] (86) at (11.25, -2) {$\multinomial[K]$};
		\node [style=black dot] (87) at (11.25, -0.75) {};
		\node [style=none] (88) at (11.25, -3.25) {};
		\node [style=none] (89) at (12.75, 0.75) {};
		\node [style=medium box] (90) at (9.75, 0.75) {$\arr$};
		\node [style=none] (91) at (7, -0.75) {$=$};
	\end{pgfonlayer}
	\begin{pgfonlayer}{edgelayer}
		\draw (73) to (75);
		\draw (77) to (81.center);
		\draw [in=-90, out=165] (75) to (83.center);
		\draw (83.center) to (82.center);
		\draw (85.center) to (90);
		\draw (84.center) to (89.center);
		\draw [in=165, out=-90] (90) to (87);
		\draw [in=-90, out=15] (87) to (89.center);
		\draw (87) to (86);
		\draw (86) to (88.center);
		\draw [in=-90, out=15] (75) to (77);
		\draw (74.center) to (73);
	\end{pgfonlayer}
\end{tikzpicture}}}
\end{equation}
\end{theorem}

\begin{myproof}
Accumulation $\acc$ and arragement $\arr$ are the retraction and
section part of the split idempotent $\tupperm \coloneqq \arr \klafter
\acc$ described in
Lemma~\ref{ArrLem}. Lemma~\ref{SplitIdempotentSuffStatLem} applies
since:
\[ \begin{array}{rcl}
\big(\tupperm \klafter \iid[K]\big)(\omega)
& = &
\displaystyle \sum_{\pi\in\Perm(\{1,\ldots,K\})} \, \sum_{\vec{x}\in X^{K}} \,
   \frac{\omega^{K}(\vec{x})}{K!} \,\Bigket{x_{\pi(1)}, \ldots, x_{\pi(K)}}
\\[+1.2em]
& = &
\displaystyle \sum_{\vec{x}\in X^{K}} \, \left(\sum_{\pi\in\Perm(\{1,\ldots,K\})} \, 
   \frac{\omega^{K}(x_{\pi(1)}, \ldots, x_{\pi(K)})}{K!}\right)\bigket{\vec{x}}
\\[+1.2em]
& = &
\displaystyle \sum_{\vec{x}\in X^{K}} \, \omega^{K}(\vec{x})\, \bigket{\vec{x}}
\hspace*{\arraycolsep}=\hspace*{\arraycolsep}
\omega^{K}
\hspace*{\arraycolsep}=\hspace*{\arraycolsep}
\iid[K](\omega).
\end{array} \]

\noindent The formulation in~\eqref{AccSuffStatEqn} now follows
because, by definition, $\multinomial[K] = \acc \klafter \iid[K]$,
see~\eqref{MultinomialEqn}. \QED
\end{myproof}

We think that Equation~\eqref{AccSuffStatEqn} expresses a fundamental
and elementary relationship between the combinatorial and
probabilistic properties of sequences and multisets, in terms of the
idd and multinomial distributions.  Although these distributions
appear frequently in the literature, this (diagrammatic)
equation~\eqref{AccSuffStatEqn} has not been identified before. There
is a single source --- as far as we are aware ---
namely~\cite[\S2.2]{Bishop06}, where it is mentioned that a particular
sum --- amounting to accumulation --- is sufficient as statistic for
iid. But the situation is not elaborated further and arrangement
does not occur, like in~\eqref{AccSuffStatEqn}.

The reference~\cite[Thm.~4]{Jacobs21b} contains the following two
equations.
\[ \begin{array}{rclcrcl}
\arr \klafter \multinomial[K]
& = &
\iid[K]
& \qquad\mbox{and}\qquad &
\acc \klafter \iid[K]
& = &
\multinomial[K].
\end{array} \]

\noindent Both can be obtained from the above theorem by discarding
one of the outgoing wires in~\eqref{AccSuffStatEqn}. However,
Equation~\eqref{AccSuffStatEqn} expresses more than these two
equations obtained by marginalisation, since it involves a joint state
formulation.

\auxproof{
We show how the accumulation function $\acc \colon X^{K} \rightarrow
\natMlt[K](X)$, given by $\acc(\vec{x}) = \sum_{i}1\ket{x_i}$, is a
sufficient statistic for the channel $\iid[K] \colon \Dst(X) \chanto
X^{K}$ of independent and identically distributed elements of $X$.  We
include a short proof, following the
formulation~\eqref{SuffStatKetEqn}. For $\omega\in\Dst(X)$,
\[ \begin{array}{rcl}
\tuple{\idmap,\acc} \push \iid[K](\omega)
%\hspace*{\arraycolsep}=\hspace*{\arraycolsep}
%\Dst\big(\tuple{\idmap,\acc}\big)\Big(\iid[K](\omega)\Big)
& = &
\displaystyle\sum_{\vec{x}\in X^{K}} \omega^{K}(\vec{x})
   \, \bigket{\vec{x}, \acc(\vec{x})}
\\[+0.5em]
& = &
\displaystyle\sum_{\varphi\in\natMlt[K](X)} \, \sum_{\vec{x}\in\acc^{-1}(\varphi)} 
   \coefm{\varphi} \cdot\textstyle{\displaystyle\prod}_{y}\, 
   \omega(x)^{\acc(\vec{x})(y)} \cdot \displaystyle  \frac{1}{\coefm{\varphi}}
   \, \bigket{\vec{x}, \varphi}
\\[+1.4em]
& = &
\displaystyle\sum_{\varphi\in\natMlt[K](X)} \, \sum_{\vec{x}\in\acc^{-1}(\varphi)} 
   \multinomial[K](\omega)(\varphi) \cdot \arr(\varphi)(\vec{x})
   \, \bigket{\vec{x}, \varphi}
\\[+0.8em]
& \smash{\stackrel{\eqref{TupleKetEqn}}{=}} &
\tuple{\arr, \idmap} \push \multinomial[K](\omega).
\end{array} \]
}

We elaborate in the concrete situation of Theorem~\ref{AccSuffStatThm}
on two aspects discussed in Remark~\ref{SuffStatRem}.  First, the
sufficiency of accumulation can also be expressed (and proven
directly) via arbitrary predicates $p\in\Pred\big(X^{K}\big)$ and
$q\in\Pred\big(\Mlt[K](X)\big)$, in the style
of~\eqref{SuffStatPredEqn}, as an adjointness property:
\[ \begin{array}{rcl}
\iid[K](\omega) \models p\andthen (\acc \pull q)
& = &
\displaystyle\sum_{\vec{x}\in X^{K}} \omega^{K}(\vec{x}) \cdot p\big(\vec{x}\big)
   \cdot q\big(\acc(\vec{x})\big)
\\[-0.6em]
& = &
\displaystyle\sum_{\varphi\in\Mlt[K](X)} 
   \coefm{\varphi} \cdot\textstyle{\displaystyle\prod}_{y}\, 
   \omega(x)^{\acc(\vec{x})(y)} \cdot \displaystyle 
   \left(\sum_{\vec{x}\in\acc^{-1}(\varphi)} 
   \frac{1}{\coefm{\varphi}} \cdot p\big(\vec{x}\big)\right)
   \cdot q(\varphi)
\\[+1.0em]
& = &
\displaystyle\sum_{\varphi\in\Mlt[K](X)} 
   \multinomial[K](\omega)(\varphi) \cdot \big(\arr \pull p\big)(\varphi)
   \cdot q(\varphi)
\\[+0.8em]
& = &
\multinomial[K](\omega) \models (\arr \pull p) \andthen q.
\end{array} \]

\noindent Next, we illustrate how in this case the dependency on
$\omega$ disappears in iid if we condition on accumulation (to a fixed
multiset $\varphi$), as in~\eqref{SuffStatDaggerEqn}:
\[ \begin{array}{rcl}
\iid[K](\omega)\big|_{\acc \pull \indic{\varphi}}
\hspace*{\arraycolsep}\smash{\stackrel{\eqref{SuffStatDaggerEqn}}{=}}\hspace*{\arraycolsep}
\displaystyle\sum_{\vec{x}\in\acc^{-1}(\varphi)} \frac{\omega^{K}(\vec{x})}
   {\Dst(\acc)(\iid[K](\omega))(\varphi)}\,\bigket{\vec{x}}
& \smash{\stackrel{\eqref{MultinomialEqn}}{=}} &
\displaystyle\sum_{\vec{x}\in\acc^{-1}(\varphi)} 
   \frac{\prod_{y}\, \omega(y)^{\acc(\vec{x})(y)}}
   {\multinomial[K](\omega)(\varphi)}\,\bigket{\vec{x}}
\\
& = &
\displaystyle\sum_{\vec{x}\in\acc^{-1}(\varphi)} 
   \frac{\prod_{y}\, \omega(y)^{\varphi(y)}}
   {\coefm{\varphi} \cdot \prod_{y}\, \omega(y)^{\varphi(y)}}\,\bigket{\vec{x}}
\hspace*{\arraycolsep}\smash{\stackrel{\eqref{ArrEqn}}{=}}\hspace*{\arraycolsep}
\arr(\varphi).
\end{array} \]

\section{Multiplicity count as sufficient statistic}\label{MulcountSuffStatSec}

Accumulation abstracts away from the order of elements in a sequence,
by turning the sequence into a multiset. Multiplicity count $\mulcount
\colon \Mlt[K](X) \rightarrow \MP(K)$ is a function that abstracts
away from the elements themselves, and only looks at their
multiplicities. Recall, $\mulcount(\varphi) =
\sum_{x\in\supp(\varphi)} 1\bigket{\varphi(x)}$. The question arises:
is multiplicity count also a sufficient statistic, and if so, for
which channel? This section provides an answer, based on a split
idempotent, like in Lemma~\ref{SplitIdempotentSuffStatLem}.  Indeed,
in Proposition~\ref{MulcountProp} we have seen that multiplicity count
$\mulcount$ is the retraction part of the split idempotent $\eltperm$
of element permutations, given by $\eltperm = \stack \klafter
\mulcount$.

We first introduce a \emph{swap} version of multinomial distribution,
defined for a distribution $\omega\in\Dst(X)$ on a finite set $X$ as:
\begin{equation}
\label{SwapMultinomialEqn}
\begin{array}{rcl}
\swapmultinomial[K](\omega)
& \coloneqq &
\displaystyle \sum_{\pi\in\Perm(X)}\,
   \frac{\multinomial[K]\big(\Dst(\pi)(\omega)\big)}{\setsize{X}!},
\end{array}
\end{equation}

\noindent For instance, using the same distribution
$\frac{1}{8}\ket{a} + \frac{1}{2}\ket{b} + \frac{3}{8}\ket{c}$ as in
the illustration of the multinomial in
Subsection~\ref{DistributionSubsec}, we now get:
\[ \begin{array}{rcl}
\lefteqn{\textstyle\swapmultinomial[3]\Big(\frac{1}{8}\ket{a} + 
   \frac{1}{2}\ket{b} + \frac{3}{8}\ket{c}\Big)}
\\[+0.4em]
& = &
\frac{23}{384}\Bigket{3\ket{a}} + 
   \frac{29}{256}\Bigket{2\ket{a} + 1\ket{b}} + 
   \frac{29}{256}\Bigket{1\ket{a} + 2\ket{b}} + 
   \frac{23}{384}\Bigket{3\ket{b}} + 
   \frac{29}{256}\Bigket{2\ket{a} + 1\ket{c}} \, + 
\\[+0.4em]
& & \;\; 
   \frac{9}{64}\Bigket{1\ket{a} + 1\ket{b} + 1\ket{c}} + 
   \frac{29}{256}\Bigket{2\ket{b} + 1\ket{c}} + 
   \frac{29}{256}\Bigket{1\ket{a} + 2\ket{c}} + 
   \frac{29}{256}\Bigket{1\ket{b} + 2\ket{c}} + 
   \frac{23}{384}\Bigket{3\ket{c}}
\end{array} \]

%% 23/384|3|a>> + 
%% 29/256|2|a> + 1|b>> + 
%% 29/256|1|a> + 2|b>> + 
%% 23/384|3|b>> + 
%% 29/256|2|a> + 1|c>> + 
%% 9/64|1|a> + 1|b> + 1|c>> + 
%% 29/256|2|b> + 1|c>> + 
%% 29/256|1|a> + 2|c>> + 
%% 29/256|1|b> + 2|c>> + 
%% 23/384|3|c>>

% S = Space('a', 'b', 'c')
% print( SwapMultinomial(3)(DState([Frac(1,8), Frac(1,2), Frac(3,8)], S)) )

\noindent We see that multisets with the same multiplicity count have
the same probability. This happens since the elements in $X$ do not
really play a role in this definition. Hence, really, we should not be
using distributions, but `divisions', as element-free distributions,
see~\cite{Jacobs22c}. This is left for future work.

\begin{theorem}
\label{MulcountSuffStatThm}
Let $X$ be a finite set and let $K \leq \setsize{X}$.  Multiplicity
count is a sufficient statistic for the swapped multinomial, via the
stack channel $\stack$, as expressed by the following sufficiency
equation between channels $\Dst(X) \chanto \Mlt[K](X)\times\MP(K)$.
\begin{equation}
\label{MulcountSuffStatEqn}
\vcenter{\hbox{\begin{tikzpicture}
	\begin{pgfonlayer}{nodelayer}
		\node [style=medium box] (73) at (2.5, -2) {$\swapmultinomial[K]$};
		\node [style=none] (74) at (2.5, -3.25) {};
		\node [style=black dot] (75) at (2.5, -0.75) {};
		\node [style=medium box] (77) at (4, 0.75) {$\mulcount$};
		\node [style=none] (81) at (4, 1.75) {};
		\node [style=none] (82) at (1, 1.75) {};
		\node [style=none] (83) at (1, 0.75) {};
		\node [style=none] (84) at (12.75, 1.75) {};
		\node [style=none] (85) at (9.75, 1.75) {};
		\node [style=medium box] (86) at (11.25, -2) {$\partmultinomial[K]$};
		\node [style=black dot] (87) at (11.25, -0.75) {};
		\node [style=none] (88) at (11.25, -3.25) {};
		\node [style=none] (89) at (12.75, 0.75) {};
		\node [style=medium box] (90) at (9.75, 0.75) {$\stack$};
		\node [style=none] (91) at (7, -0.75) {$=$};
	\end{pgfonlayer}
	\begin{pgfonlayer}{edgelayer}
		\draw (73) to (75);
		\draw (77) to (81.center);
		\draw [in=-90, out=165] (75) to (83.center);
		\draw (83.center) to (82.center);
		\draw (85.center) to (90);
		\draw (84.center) to (89.center);
		\draw [in=165, out=-90] (90) to (87);
		\draw [in=-90, out=15] (87) to (89.center);
		\draw (87) to (86);
		\draw (86) to (88.center);
		\draw [in=-90, out=15] (75) to (77);
		\draw (74.center) to (73);
	\end{pgfonlayer}
\end{tikzpicture}}}
\end{equation}

\noindent It involves the `partition multinomial' channel
$\partmultinomial[K] \colon \Dst(X) \chanto \MP(K)$ satisfying:
\begin{equation}
\label{PartMultinomialEqn}
\begin{array}{rcccccl}
\partmultinomial[K] 
& \coloneqq &
\mulcount \klafter \swapmultinomial[K]
& = &
\mulcount \klafter \multinomial[K]
& = &
\mulcount \klafter \acc \klafter \iid[K].
\end{array}
\end{equation}
\end{theorem}

\begin{myproof}
We use Lemma~\ref{SplitIdempotentSuffStatLem} with split idempotent
$\eltperm = \stack \klafter \acc$. We thus have to prove that
$\eltperm \klafter \swapmultinomial[K] = \swapmultinomial[K]$. This
follows from the fact that $\swapmultinomial[K] = \eltperm \klafter
\multinomial[K]$. The proof uses that the multinomial channel
$\multinomial[K]$ forms a natural transformation $\Dst \Rightarrow
\Dst\Mlt[K]$, see~\cite{Jacobs21b}.
\[ \begin{array}{rcl}
\big(\eltperm \klafter \multinomial[K]\big)(\omega)
& = &
\displaystyle\sum_{\pi\in\Perm(X)} \, \sum_{\varphi\in\Mlt[K](X)} \,
   \frac{\multinomial[K](\omega)(\varphi)}{\setsize{X}!} 
   \, \Bigket{\Mlt(\pi)(\varphi)}
   \qquad \mbox{see Proposition~\ref{MulcountProp}~\eqref{MulcountPropPermute}}
\\[+1.4em]
& = &
\displaystyle\sum_{\pi\in\Perm(X)} \, 
   \frac{\Dst\Mlt(\pi)(\multinomial[K](\omega))}{\setsize{X}!} 
\\[+1em]
& = &
\displaystyle\sum_{\pi\in\Perm(X)} \, 
   \frac{\multinomial[K](\Dst(\pi)(\omega))}{\setsize{X}!} 
\hspace*{\arraycolsep}\smash{\stackrel{\eqref{SwapMultinomialEqn}}{=}}\hspace*{\arraycolsep}
\swapmultinomial[K](\omega).
\end{array} \]

\noindent Lemma~\ref{SplitIdempotentSuffStatLem} now provides a
sufficient statistic situation with channel:
\[ \begin{array}{rcccccl}
\mulcount \klafter \swapmultinomial[K]
& = &
\mulcount \klafter \eltperm \klafter \multinomial[K]
& = &
\mulcount \klafter \stack \klafter \mulcount \klafter \acc \klafter \iid[K]
& = &
\mulcount \klafter \acc \klafter \iid[K].
\end{array} \]

\noindent This composite is what we have called the partition
multinomial $\partmultinomial[K]$ in~\eqref{PartMultinomialEqn}. \QED
\end{myproof}

We like to get a better handle on this partition multinomial
$\partmultinomial[K]$.

\begin{proposition}
\label{PartMultinomialProp}
Let $X$ be a finite set and let $K$ be a natural number with $1 \leq
K\leq \setsize{X}$.
\begin{enumerate}
\item \label{PartMultinomialPropCoef} For $\sigma\in\MP(K)$ define a
  \emph{multinomial coefficient for partitions} $\partcoefm{\sigma}
  \in \NNO$ as:
\[ \begin{array}{rcl}
\partcoefm{\sigma}
& \coloneqq &
\displaystyle\frac{K!}{\prod_{1\leq i\leq K}\, (i!)^{\sigma(i)}}.
\end{array} \]

\noindent Then, for each $\varphi\in\Mlt[K](X)$,
\[ \begin{array}{rclcrcl}
\facto{\varphi}
& = &
\displaystyle\prod_{1\leq i \leq K} \big(i!\big)^{\mulcount(\varphi)(i)}
& \qquad\mbox{and thus}\qquad &
\coefm{\varphi}
& = &
\partcoefm{\mulcount(\varphi)}.
\end{array} \]

\item \label{PartMultinomialPropPart} The partition multinomial can be
  described concretely, on $\omega \in\Dst(X)$, as:
\begin{equation}
\label{PartMultinomialConcreteEqn}
\begin{array}{rcl}
\partmultinomial[K](\omega)
& = &
\displaystyle\sum_{\sigma\in\MP(K)} \partcoefm{\sigma} \cdot
   \sum_{\varphi\in\mulcount^{-1}(\sigma)} \, 
   \prod_{x\in X}\, \omega(x)^{\varphi(x)} \, \bigket{\sigma}.
\end{array}
\end{equation}
\end{enumerate}
\end{proposition}

\begin{myproof}
\begin{enumerate}
\item We first prove the equation $\facto{\varphi} = \prod_{1\leq i
  \leq K} \big(i!\big)^{\mulcount(\varphi)(i)}$, for
  $\varphi\in\Mlt[K](X)$, by induction on $K\geq 1$. The case $K = 1$
  is trivial, since then $\varphi = 1\ket{x}$, for some $x\in X$, so
  that $\mulcount(\varphi) = 1\ket{1}$. Then: $\facto{\varphi} = 1 =
  (1!)^{1}$. For the induction step, write
  $\varphi\in\Mlt[K\!+\!1](X)$ as $\varphi = \psi + 1\ket{x}$, for
  some $x\in\supp(\varphi)$. Then $\psi\in\Mlt[K](X)$.
\[ \begin{array}{rcl}
\displaystyle\prod_{1\leq i \leq K+1}\! \big(i!\big)^{\mulcount(\varphi)(i)}
& = &
\displaystyle \big((K\!+\!1)!\big)^{\mulcount(\psi+1\ket{x})(K+1)} \cdot
   \prod_{1\leq i \leq K} \big(i!\big)^{\mulcount(\psi+1\ket{x})(i)}
\\[+1em]
& = &
\begin{cases}
(K\!+\!1)! & \mbox{if }\psi = K\ket{x}
\\
\displaystyle\prod_{1\leq i \leq K} \big(i!\big)^{\mulcount(\psi)(i)} \cdot
   (\psi(x)\!+\!1)  & \mbox{otherwise}
\end{cases}
\\[+1.7em]
& \smash{\stackrel{\text{(IH)}}{=}} &
\left\{\begin{array}{ll}
\facto{(\psi+1\ket{x})} & \mbox{if }\psi = K\ket{x}
\\
\facto{\psi} \cdot (\psi(x)\!+\!1) & \mbox{otherwise}
\end{array}\right\}
\hspace*{\arraycolsep}=\hspace*{\arraycolsep}
\facto{(\psi+1\ket{x})}
\hspace*{\arraycolsep}=\hspace*{\arraycolsep}
\facto{\varphi}.
\end{array} \]

\noindent For the second equation fix $\varphi\in\Mlt[K](X)$. Then, by
what we have just shown:
\[ \begin{array}{rcccccl}
\coefm{\varphi}
& = &
\displaystyle\frac{K!}{\facto{\varphi}}
& = &
\displaystyle\frac{K!}{\prod_{1\leq i \leq K} \big(i!\big)^{\mulcount(\varphi)(i)}}
& = &
\partcoefm{\mulcount(\varphi)}.
\end{array} \]

\item By the previous item:
\[ \begin{array}[b]{rcl}
\partmultinomial[K](\omega)
\hspace*{\arraycolsep}\smash{\stackrel{\eqref{PartMultinomialEqn}}{=}}\hspace*{\arraycolsep}
\Dst\big(\mulcount\big)\Big(\multinomial[K](\omega)\Big)
& \smash{\stackrel{\eqref{MultinomialEqn}}{=}} &
\displaystyle\sum_{\varphi\in\Mlt[K](X)} \coefm{\varphi} \cdot
   \prod_{x\in X} \omega(x)^{\varphi(x)} \, \bigket{\mulcount(\varphi)}
\\[+1em]
& = &
\displaystyle\sum_{\sigma\in\MP(K)} 
   \left(\sum_{\varphi\in\mulcount^{-1}(\sigma)} \coefm{\varphi} \cdot
   \prod_{x\in X} \omega(x)^{\varphi(x)}\right) \, \bigket{\sigma}
\\[+1em]
& = &
\displaystyle\sum_{\sigma\in\MP(K)} \partcoefm{\sigma} \cdot
   \left(\sum_{\varphi\in\mulcount^{-1}(\sigma)} \,
   \prod_{x\in X} \omega(x)^{\varphi(x)}\right) \, \bigket{\sigma}.
\end{array} \eqno{\QEDbox} \]
\end{enumerate}
\end{myproof}

We conclude with an illustration:
\[ \begin{array}{rcl}
\partmultinomial[3]\Big(\frac{1}{8}\ket{a} + 
   \frac{1}{2}\ket{b} + \frac{3}{8}\ket{c}\Big)
& = &
\frac{9}{64}\Bigket{3\ket{1}} + 
   \frac{87}{128}\Bigket{1\ket{1} + 1\ket{2}} + 
   \frac{23}{128}\Bigket{1\ket{3}}.
\end{array} \]

\noindent This channel $\partmultinomial[K]$ acts like a (Kingman) paintbox,
see~\cite{BroderickPJ13}.

\section{Size of partitions as sufficient statistic}\label{SizeSuffStatSec}

We now write $\size(\sigma) \coloneqq \|\sigma\|$, for
$\sigma\in\MP(K)$, giving a function $\size \colon \MP(K) \rightarrow
\{1,\ldots,K\}$. These sizes vary, for a fixed sum,
see~\eqref{PartitionsFour}. We also introduce notation for the product
of the numbers in a partition: $\maal(\sigma) \coloneqq \prod_{i} \,
i^{\sigma(i)}$. For instance, the partitions in~\eqref{PartitionsFour}
have product, in order: $1$, $2$, $3$, $4$, $4$.

It is known from~\cite{Ewens72} that the $\size$ function is a
sufficient statistic, namely for the channel of Ewens distributions,
see also~\cite[\S2.2]{Crane16} or~\cite{Joyce98}. We elaborate this
situation in terms of Lemma~\ref{DetDaggerLem} --- which leads to
split dagger idempotents.  Due to space constraints, we only look at
the main lines here. The appendix contains more details.

The Ewens distribution, from~\cite{Ewens72}, can be described as a channel
$\ewens[K] \colon \pR \rightarrow \Dst\big(\MP(K)\big)$, given by:
\begin{equation}
\label{EwensDstEqn}
\begin{array}{rcl}
\ewens[K](t)
& \coloneqq &
\displaystyle \frac{K!}{\prod_{0\leq i < K} t \!+\! i} \cdot \sum_{\sigma\in\MP(K)} \,
   \frac{t^{\size(\sigma)}}{\facto{\sigma} \cdot \maal(\sigma)} 
   \, \bigket{\sigma}.
\end{array}
\end{equation}

\noindent We show that by updating with a fixed size $n$, where $1\leq
n\leq K$, the dependence on the parameter $t$ disappears, and gives
us our dagger $\size^{\dag}$.
\[ \begin{array}{rcl}
\ewens[K](t)\big|_{\size\pull\indic{n}}
& = &
\displaystyle\sum_{\sigma\in\size^{-1}(n)} 
   \frac{\frac{t^{\size(\sigma)}}{\sfacto{\sigma} \cdot \maal(\sigma)}}
        {\sum_{\tau\in\size^{-1}(n)} \frac{t^{\size(\tau)}}
            {\sfacto{\tau} \cdot \maal(\tau)}}\,\bigket{\sigma}
\\[+1.5em]
& = &
\displaystyle\sum_{\sigma\in\size^{-1}(n)} 
   \frac{\frac{1}{\sfacto{\sigma} \cdot \maal(\sigma)}}
        {\sum_{\tau\in\size^{-1}(n)} \frac{1}
            {\sfacto{\tau} \cdot \maal(\tau)}}\,\bigket{\sigma}
\hspace*{\arraycolsep}\eqqcolon\hspace*{\arraycolsep}\size^{\dag}(n).
\end{array} \]

\begin{theorem}
\label{SizeSuffStatThm}
For $K\geq 1$, the size function $\size \colon \MP(K) \rightarrow
\{1,\ldots,K\}$ is a sufficient statistic for the Ewens
channel~\eqref{EwensDstEqn}, via the equality of channels $\pR \chanto
\MP(K) \times \{1,\ldots,K\}$ in:
\[ \vcenter{\hbox{\begin{tikzpicture}
	\begin{pgfonlayer}{nodelayer}
		\node [style=medium box] (73) at (3.5, -2) {$\ewens[K]$};
		\node [style=none] (74) at (3.5, -3.25) {};
		\node [style=black dot] (75) at (3.5, -0.75) {};
		\node [style=medium box] (77) at (5, 0.75) {$\size$};
		\node [style=none] (81) at (5, 1.75) {};
		\node [style=none] (82) at (2, 1.75) {};
		\node [style=none] (83) at (2, 0.75) {};
		\node [style=none] (84) at (12, 1.75) {};
		\node [style=none] (85) at (9, 1.75) {};
		\node [style=medium box] (86) at (10.5, -2) {$\stirling[K]$};
		\node [style=black dot] (87) at (10.5, -0.75) {};
		\node [style=none] (88) at (10.5, -3.25) {};
		\node [style=none] (89) at (12, 0.75) {};
		\node [style=medium box] (90) at (9, 0.75) {$\size^{\dag}$};
		\node [style=none] (91) at (7, -0.75) {$=$};
	\end{pgfonlayer}
	\begin{pgfonlayer}{edgelayer}
		\draw (73) to (75);
		\draw (77) to (81.center);
		\draw [in=-90, out=165] (75) to (83.center);
		\draw (83.center) to (82.center);
		\draw (85.center) to (90);
		\draw (84.center) to (89.center);
		\draw [in=165, out=-90] (90) to (87);
		\draw [in=-90, out=15] (87) to (89.center);
		\draw (87) to (86);
		\draw (86) to (88.center);
		\draw [in=-90, out=15] (75) to (77);
		\draw (74.center) to (73);
	\end{pgfonlayer}
\end{tikzpicture}}}
\qquad\mbox{where}\qquad
{\begin{array}{l}
\text{$\stirling[K] \colon \pR \rightarrow \Dst\big(\{1,\ldots,K\}\big)$ is the}
\\[-0.2em]
\text{`Stirling' channel given by:}
\\
\quad\begin{array}{rcl}
\lefteqn{\stirling[K](t)}
\\[-0.5em]
& \coloneqq &
\displaystyle \frac{1}{\prod_{0\leq i < K} t \!+\! i} \cdot \sum_{1\leq k\leq K} 
   \left[\begin{matrix} K \\ k \end{matrix}\right]\cdot t^{k} \bigket{k}.
\end{array}
\end{array}} \]

\noindent The latter \emph{Stirling} distribution $\stirling[K]$
involves the Stirling numbers of the first kind
$\left[\begin{smallmatrix} K \\ k \end{smallmatrix}\right]$,
see~\cite{Jacobs22f} for more details. It is sometimes called the
Chinese restaurant table distribution. \QED
\end{theorem}

\section{Sums of sequences as sufficient statistic}\label{SumSuffStatSec}

For a fixed number $K$ we consider the addition function $\som \colon
\NNO^{K} \rightarrow \NNO$. It is the retraction part of a split
idempotent, with section $\som^{\dag} \colon \NNO \rightarrow
\Dst\big(\NNO^{K}\big)$ given by:
\[ \begin{array}{rcccl}
\som^{\dag}(n)
& \coloneqq &
\displaystyle\sum_{\vec{k}\in\som^{-1}(n)} \, 
   \frac{\binom{n}{\vec{k}}}{K^{n}}\,\bigket{\vec{k}}
& = &
\displaystyle\sum_{\vec{k}\in\som^{-1}(n)} \, 
   \frac{n!}{K^{n}\cdot \prod_{i} k_{i}!}\,\bigket{\vec{k}}.
\end{array} \]

\noindent This yields a distribution by
Lemma~\ref{AccLem}~\eqref{AccLemSum} since a sequence $\vec{k}\in\NNO$
with $\som(\vec{k}) = n$ can be identified with a multiset in
$\Mlt[n](\{1,\ldots,K\})$. Our final result is a well known instance
of sufficiency, involving the Poisson channel $\poissonname \colon \pR
\rightarrow \infDst(\NNO)$ given by $\poisson[\lambda] \coloneqq
\sum_{n}\, e^{-\lambda}\cdot \frac{\lambda^{n}}{n!}\bigket{n}$, with
infinite support. The appendix contains a proof.

\begin{theorem}
\label{SumSuffStatThm}
For $K\geq 1$, the addition function $\som \colon \NNO^{K} \rightarrow
\NNO$ is a sufficient statistic for $K$-fold possion product
$\bigotimes_{K}\poissonname \colon \pR \rightarrow
\Dst\big(\NNO^{K}\big)$ in the following equality of channels $\pR
\chanto \NNO^{K} \times \NNO$.
\[ \vcenter{\hbox{\begin{tikzpicture}
	\begin{pgfonlayer}{nodelayer}
		\node [style=medium box] (73) at (2.5, -2) {$\bigotimes_{K}\poissonname$};
		\node [style=none] (74) at (2.5, -3.25) {};
		\node [style=black dot] (75) at (2.5, -0.75) {};
		\node [style=medium box] (77) at (4, 0.75) {$\som$};
		\node [style=none] (81) at (4, 1.75) {};
		\node [style=none] (82) at (1, 1.75) {};
		\node [style=none] (83) at (1, 0.75) {};
		\node [style=none] (84) at (12.75, 1.75) {};
		\node [style=none] (85) at (9.75, 1.75) {};
		\node [style=medium box] (86) at (11.25, -2) {$\poisson[K-]$};
		\node [style=black dot] (87) at (11.25, -0.75) {};
		\node [style=none] (88) at (11.25, -3.25) {};
		\node [style=none] (89) at (12.75, 0.75) {};
		\node [style=medium box] (90) at (9.75, 0.75) {$\som^{\dag}$};
		\node [style=none] (91) at (7, -0.75) {$=$};
	\end{pgfonlayer}
	\begin{pgfonlayer}{edgelayer}
		\draw (73) to (75);
		\draw (77) to (81.center);
		\draw [in=-90, out=165] (75) to (83.center);
		\draw (83.center) to (82.center);
		\draw (85.center) to (90);
		\draw (84.center) to (89.center);
		\draw [in=165, out=-90] (90) to (87);
		\draw [in=-90, out=15] (87) to (89.center);
		\draw (87) to (86);
		\draw (86) to (88.center);
		\draw [in=-90, out=15] (75) to (77);
		\draw (74.center) to (73);
	\end{pgfonlayer}
\end{tikzpicture}}}
\qquad\mbox{where}\qquad
{\left\{\begin{array}{rcl}
\big(\textstyle\bigotimes_{K}\poissonname\big)(\lambda)
& = &
\poisson[\lambda] \otimes \cdots \otimes \poisson[\lambda]
\\[+0.2em]
\poisson[K-](\lambda)
& = &
\poisson[K\lambda].
\end{array}\right.} \eqno{\QEDbox} \]
\end{theorem}

An analogous example uses geometric instead of Poisson distributions.

\section{Conclusions}\label{ConclusionSec}

This paper elaborates several examples of a sufficient statistic, in
the recently introduced `adjoint' formulation using string diagrams
of~\cite{Fritz20}. It does so for discrete distributions, since
sufficiency is not well-developed there --- in contrast to the
continuous case. Four examples are described, following the same
pattern, by identifying the relevant split idempotents --- that arise
via the Fisher-Neyman factorisation theorem. The paper's focus is on
concrete mathematical (combinatorial) structure, and not so much on
theory development. Ewens' distributions capture probabilistic
mutations for multiset partitions. In~\cite{Jacobs22g} this approach
is extended to other data types --- including set partitions, also
studied in~\cite{JacobsS23a} --- and leads to many more examples of
sufficient statistics. There is room for further investigations of,
for instance. the use of `divisions' from~\cite{Jacobs22c} instead of
the `partition' distributions in Section~\ref{MulcountSuffStatSec}.

%\bibliography{/home/bart/svn/bart/Tex/bib}

\appendix
\section{Appendix}

\subsection*{Proof of Theorem~\ref{SizeSuffStatThm}}

For $K\geq 1$ we have the Ewens and Stirling channels:
\[ \xymatrix@C-0.3pc{
\pR\ar[rr]|-{\circ}^-{\ewens[K]} & & \MP(K)
& &
\pR\ar[rr]|-{\circ}^-{\stirling[K]} & & \{1,\ldots,K\}
} \]

\noindent For convenience, we repeat the definitions from
Section~\ref{SizeSuffStatSec}: for a reproduction parameter $t\in\pR$,
\begin{equation}
\label{EwensStirlingEqn}
\begin{array}{rcl}
\ewens[K](t)
& = &
\displaystyle \frac{K!}{\prod_{0\leq i < K} t \!+\! i} \cdot \sum_{\sigma\in\MP(K)} \,
   \frac{t^{\size(\sigma)}}{\facto{\sigma} \cdot \maal(\sigma)} 
   \, \bigket{\sigma}
\\
\stirling[K](t)
& = &
\displaystyle \frac{1}{\prod_{0\leq i < K} t \!+\! i} \cdot \sum_{1\leq k\leq K} 
   \left[\begin{matrix} K \\ k \end{matrix}\right]\cdot t^{k} \bigket{k}.
\end{array}
\end{equation}

In the description of the Stirling distribution $\stirling[K](t)$ We
write $\left[\begin{smallmatrix} K \\ k \end{smallmatrix}\right]$ for
the Stirling number of the first kind. These numbers are determined by
the equations:
\begin{equation}
\label{StirlingEqn}
\begin{array}{rclcrccclcrcl}
\left[\begin{matrix} 0 \\ 0 \end{matrix}\right]
& = &
1
& \qquad &
\left[\begin{matrix} 0 \\ k \end{matrix}\right]
& = &
\left[\begin{matrix} k \\ 0 \end{matrix}\right]
& = &
0
& \qquad &
\left[\begin{matrix} n\!+\!1 \\ k \end{matrix}\right]
& = &
n\cdot \left[\begin{matrix} n \\ k \end{matrix}\right]
+
\left[\begin{matrix} n \\ k\!-\!1 \end{matrix}\right].
\end{array}
\end{equation}

\noindent One can then derive, by induction on $K\geq 1$, that for all
$t>0$,
\begin{equation}
\label{StirlingConvexEqn}
\begin{array}{rcl}
\displaystyle\sum_{1\leq k \leq K}
\left[\begin{matrix} K \\ k \end{matrix}\right]\cdot t^{k}
& = &
\displaystyle \prod_{0\leq i < K} t \!+\! i
\end{array} 
\end{equation}

\noindent Therefore, the probabilities in the above Stirling
distributions add up to one.

\auxproof{
For instance:
\[ \begin{array}{rcl}
\stirling[4](1)
& = &
\frac{1}{24}\bigket{1} + \frac{11}{24}\bigket{2} + 
   \frac{1}{4}\bigket{3} + \frac{1}{4}\bigket{4} 
\\[+0.3em]
\stirling[5](2)
& = &
\frac{1}{15}\bigket{1} + \frac{5}{18}\bigket{2} + \frac{7}{18}\bigket{3} + 
   \frac{2}{9}\bigket{4} + \frac{2}{45}\bigket{5}. 
\end{array} \]
}

We use the following general result about the Ewens and Stirling
distributions~\eqref{EwensStirlingEqn}. The part about Ewens
is copied from~\cite{Jacobs22b}.

\begin{lemma}
\label{EwensStirlingLem}
Fix $K\in\NNO$ and $t\in\pR$.
\begin{enumerate}
\item \label{EwensStirlingLemCommute} There is a commuting rectangle
  of `draw-add' channels:
\[ \xymatrix@R-0.8pc{
\MP(K)\ar[d]_{\size}\ar[rr]^-{\partdrawadd[K](t)} 
   & & \Dst\big(\MP(K\!+\!1)\big)\ar[d]^-{\Dst(\size)}
\\
\{1,\ldots,K\}\ar[rr]^-{\setdrawadd[K](t)} & & 
   \Dst\big(\{1,\ldots,K\!+\!1\}\big)
} \]

\noindent In which the \emph{partition} and \emph{set} draw-add channels
$\partdrawadd$ and $\setdrawadd$ are defined by:
\[ \begin{array}{rcl}
\partdrawadd(t)(\sigma)
& \coloneqq &
\displaystyle\frac{t}{K\!+\!t}\Bigket{\sigma+1\ket{1}} +
   \sum_{1\leq k \leq K} \, \frac{\sigma(k)\cdot k}{K\!+\!t}
   \Bigket{\sigma - 1\ket{k} + 1\ket{k\!+\!1}}
\\[+1em]
\setdrawadd(t)(k)
& \coloneqq &
\displaystyle\frac{t}{K\!+\!t}\bigket{k+1} +
   \frac{K}{K\!+\!t}\bigket{k}.
\end{array} \]

\item \label{EwensStirlingLemInduct} Both the Ewens and the Stirling
  distributions can be obtained inductively via:
\noindent We follow~\cite{Jacobs22b} in the inductive description of
Ewens distributions via state transformation as:
\[ \left\{\begin{array}{rcl}
\ewens[1](t)
& = &
1\Bigket{1\ket{1}}
\\
\ewens[K\!+\!1](t)
& = &
\partdrawadd[K](t) \push \ewens[K](t)
\end{array}\right.
\qquad\qquad
\left\{\begin{array}{rcl}
\stirling[1](t)
& = &
1\bigket{1}
\\
\stirling[K\!+\!1](t)
& = &
\setdrawadd[K](t) \push \stirling[K](t).
\end{array}\right. \]

\item \label{EwensStirlingLemSize} $\begin{array}{rcl}
\Dst(\size) \after \ewens[K]
& = &
\stirling[K].
\end{array}$
\end{enumerate}
\end{lemma}

In Section~\ref{SizeSuffStatSec} we have already determined the dagger
to the size function. Hence this equation in item~\eqref{EwensStirlingLemSize}
suffices to prove Theorem~\ref{SizeSuffStatThm}.

\begin{myproof}
\begin{enumerate}
\item For $\sigma\in\MP(K)$,
\[ \begin{array}{rcl}
\big(\Dst(\size) \after \partdrawadd[K](t)\big)(\sigma)
& = &
\displaystyle\frac{t}{K\!+\!t}\Bigket{\size(\sigma+1\ket{1})} +
   \sum_{1\leq k \leq K} \, \frac{\sigma(k)\cdot k}{K\!+\!t}
   \Bigket{\size(\sigma - 1\ket{k} + 1\ket{k\!+\!1})}
\\[+1em]
& = &
\displaystyle\frac{t}{K\!+\!t}\Bigket{\size(\sigma)+1} +
    \frac{\sum_{1\leq k \leq K} \,\sigma(k)\cdot k}{K\!+\!t}
   \Bigket{\size(\sigma)}
\\[+1em]
& = &
\displaystyle\frac{t}{K\!+\!t}\Bigket{\size(\sigma)+1} +
    \frac{K}{K\!+\!t}
   \Bigket{\size(\sigma)}
\\[+0.6em]
& = &
\setdrawadd[K](t)\big(\size(\sigma)\big).
\end{array} \]

\item We leave the Ewens case to the reader and concentrate on
  Stirling. First, for $K=1$,
\[ \begin{array}{rcccccl}
\stirling[1](t)
& = &
\displaystyle \frac{1}{\prod_{0\leq i < 1} t \!+\! i} \cdot \sum_{1\leq k\leq 1} 
   \left[\begin{matrix} 1 \\ k \end{matrix}\right]\cdot t^{k} \bigket{k}
& = &
\displaystyle\frac{1}{t} \cdot 
   \left[\begin{matrix} 1 \\ 1 \end{matrix}\right]\cdot t^{1} \bigket{1}
& = &
1\ket{1}.
\end{array} \]

\noindent Next, 
\[ \begin{array}{rcl}
\lefteqn{\setdrawadd[K](t) \push \stirling[K](t)}
\\
& = &
\displaystyle\sum_{1\leq \ell \leq K+1} \; \sum_{1\leq k\leq K} \,
   \setdrawadd[K](t)(k)(\ell) \cdot \stirling[K](t)(k) 
   \, \bigket{\ell}
\\[+1em]
& = &
\displaystyle \sum_{1\leq k\leq K} \,
   \frac{t}{K+t} \cdot \stirling[K](t)(k) \, \bigket{k+1} \;+\;
   \sum_{1\leq k\leq K} \,
   \frac{K}{K+t} \cdot \stirling[K](t)(k) \, \bigket{k}
\\[+0.6em]
& \smash{\stackrel{\text{(IH)}}{=}} &
\displaystyle \sum_{1\leq k\leq K} \,
   \frac{t}{K+t} \cdot \frac{1}{\prod_{0\leq i < K} t \!+\! i} \cdot 
   \left[\begin{matrix} K \\ k \end{matrix}\right]\cdot t^{k} \, \bigket{k+1} 
   \;+\; \sum_{1\leq k\leq K} \,
   \frac{K}{K+t} \cdot \frac{1}{\prod_{0\leq i < K} t \!+\! i} \cdot 
   \left[\begin{matrix} K \\ k \end{matrix}\right]\cdot t^{k} \bigket{k}
\\[+1em]
& = &
\displaystyle \frac{1}{\prod_{i < K\!+\!1} t \!+\! i} \cdot \left(
   \sum_{1\leq k\leq K} 
   \left[\begin{matrix} K \\ k \end{matrix}\right]\cdot 
   t^{k+1} \, \bigket{k+1} \;+\;
   \sum_{1\leq k\leq K} K\cdot 
   \left[\begin{matrix} K \\ k \end{matrix}\right]\cdot t^{k} 
   \, \bigket{k}\right)
\\[+1.2em]
& = &
\displaystyle \frac{1}{\prod_{i < K\!+\!1} t \!+\! i} \cdot \left(
   K\cdot \left[\begin{matrix} K \\ 1 \end{matrix}\right] 
      \cdot t^{1} \, \bigket{1}
   + 
   \Big(K\cdot \left[\begin{matrix} K \\ 2 \end{matrix}\right] +
      \left[\begin{matrix} K \\ 1 \end{matrix}\right]\Big)
      \cdot t^{2} \, \bigket{2}  \;+\; \cdots\; + \right.
\\[+1em]
& & \hspace*{10em} \displaystyle \left.
   \Big(K\cdot \left[\begin{matrix} K \\ K \end{matrix}\right] +
      \left[\begin{matrix} K \\ K\!-\!1 \end{matrix}\right]\Big)
      \cdot t^{K} \, \bigket{K} +
  \left[\begin{matrix} K \\ K \end{matrix}\right] 
      \cdot t^{K+1} \, \bigket{K\!+\!1}\right)
\\[+1.2em]
& \smash{\stackrel{\eqref{StirlingEqn}}{=}} &
\displaystyle \frac{1}{\prod_{i < K\!+\!1} t \!+\! i} \cdot \left(
   \left[\begin{matrix} K\!+\!1 \\ 1 \end{matrix}\right] 
      \cdot t^{1} \, \bigket{1}
   + 
   \left[\begin{matrix} K\!+\!1 \\ 2 \end{matrix}\right]
      \cdot t^{2} \, \bigket{2}  \; +\; \cdots\; + \right.
\\[+1em]
& & \hspace*{10em} \displaystyle \left.
   \left[\begin{matrix} K\!+\!1 \\ K \end{matrix}\right]
      \cdot t^{K} \, \bigket{K} +
  \left[\begin{matrix} K\!+\!1 \\ K\!+\!1 \end{matrix}\right] 
      \cdot t^{K+1} \, \bigket{K\!+\!1}\right)
\\[+1.2em]
& = &
\displaystyle \frac{1}{\prod_{i < K\!+\!1} t \!+\! i} \cdot \sum_{1\leq k\leq K+1} \,
   \left[\begin{matrix} K\!+\!1 \\ k \end{matrix}\right]\cdot 
   t^{k} \, \bigket{k}
\\[+0.6em]
& = &
\stirling[K\!+\!1](t).
\end{array} \]

\item By induction on $K$, using items~\eqref{EwensStirlingLemCommute}
  and~\eqref{EwensStirlingLemInduct}. \QED
\end{enumerate}
\end{myproof}

As an aside, there is a no draw-delete channel $\{1,\ldots,K+1\}
\chanto \{1,\ldots,K\}$ that commutes with size, as in
Lemma~\ref{EwensStirlingLem}~\eqref{EwensStirlingLemCommute},
see~\cite{Jacobs22b} for more information about draw-add and
draw-delete, for multisets and partitions.

\auxproof{
Addition: there is a no draw-delete channel $\{1,\ldots,K+1\} \chanto
\{1,\ldots,K\}$ that commutes with size, as in
Lemma~\ref{EwensStirlingLem}~\eqref{EwensStirlingLemCommute}. For
instance, for $K=10$, consider the following two partitions in
$\MP(11)$, both with size $5$.
\[ \begin{array}{rclcrcl}
\sigma 
& = &
3\ket{1} + 2\ket{4}
& \qquad \qquad &
\tau 
& = &
1\ket{1} + 2\ket{2} + 2\ket{3}.
\end{array} \]

\noindent Then:
\[ \begin{array}{rcl}
\partdrawdelete(\sigma)
& = &
\frac{8}{11}\Bigket{3\ket{1} + 1\ket{3} + 1\ket{4}} +
   \frac{3}{11}\Bigket{2\ket{1} + 2\ket{4}} 
\\
\partdrawdelete(\tau)
& = &
\frac{6}{11}\Bigket{1\ket{1} + 3\ket{2} + 1\ket{3}} +
  \frac{4}{11}\Bigket{2\ket{1} + 1\ket{2} + 2\ket{3}} +
  \frac{1}{11}\Bigket{2\ket{2} + 2\ket{3}}
\end{array} \]

\noindent They yield different size-outcomes:
\[ \begin{array}{rcl}
\Dst(\size)\Big(\partdrawdelete(\sigma)\Big)
& = &
\frac{10}{11}\bigket{5} + \frac{1}{11}\bigket{4}
\\
\Dst(\size)\Big(\partdrawdelete(\tau)\Big)
& = &
\frac{8}{11}\bigket{5} + \frac{3}{11}\bigket{4}.
\end{array} \]
}

\subsection*{Proof of Theorem~\ref{SumSuffStatThm}}

For $K\geq 1$ we have two $\infDst$-channels:
\[ \xymatrix@C-0.7pc{
\pR\ar[rr]|-{\circ}^-{\bigotimes_{K}\poissonname} & & \NNO^{K}
& 
\pR\ar[rr]|-{\circ}^-{\poisson[K-]} & & \NNO
\qquad\mbox{with}\qquad
{\left\{\begin{array}{rcl}
\big(\bigotimes_{K}\poissonname\big)(\lambda)
& = &
\poisson[\lambda] \otimes \cdots \otimes \poisson[\lambda]
\\
\big(\poisson[K-]\big)(\lambda)
& = &
\poisson[K\lambda].
\end{array}\right.}
} \]

\noindent In addition, there is the $\som$ sufficient statistic with
its dagger $\som^{\dag}$ in:
\[ \xymatrix{
\NNO^{K}\ar[rr]^-{\som} & & \NNO
& 
\NNO\ar[rr]|-{\circ}^-{\som^{\dag}} & & \NNO^{K}
\qquad\mbox{with}\qquad
{\left\{\begin{array}{rcl}
\som(\vec{k})
& = &
\sum_{i}k_{i}
\\
\som^{\dag}(n)
& = &
\displaystyle\sum_{\vec{k}\in\som^{-1}(n)} \frac{\binom{n}{\vec{k}}}{K^{n}}
   \,\bigket{\vec{k}}
\end{array}\right.}
} \]

\noindent We choose to prove Theorem~\ref{SumSuffStatThm} via the
predicate formulation of~\eqref{SuffStatPredEqn}. For predicates
$p\in\Pred(\NNO^{K})$ and $q\in\Pred(\NNO)$, there is an adjoint
correspondence:
\[ \begin{array}{rcl}
\big(\bigotimes_{K} \poissonname\big)(\lambda) \models 
   p \andthen (\som \pull q)
& = &
\displaystyle\sum_{\vec{k}\in\NNO^{K}} \textstyle {\displaystyle\prod}_{i} \,
   \poisson[\lambda](k_{i}) \cdot p\big(\vec{k}\big) \cdot 
   q\big(\som(\vec{k})\big)
\\[+0.8em]
& = &
\displaystyle\sum_{\vec{k}\in\NNO^{K}} \textstyle {\displaystyle\prod}_{i} \,
   \displaystyle e^{-\lambda} \cdot \frac{\lambda^{k_i}}{k_{i}!} \cdot 
   p\big(\vec{k}\big) \cdot q\big(\som(\vec{k})\big)
\\[+0.8em]
& = &
\displaystyle\sum_{\vec{k}\in\NNO^{K}} e^{-K\lambda} \cdot 
   \frac{\lambda^{\som(\vec{k})}}{\prod_{i} \, k_{i}!} \cdot 
   p\big(\vec{k}\big) \cdot q\big(\som(\vec{k})\big)
\\[+0.8em]
& = &
\displaystyle \sum_{n\in\NNO}\;\, \sum_{\vec{k}\in\som^{-1}(n)} e^{-K\lambda} \cdot 
   \frac{\lambda^{n}}{n!} \cdot \binom{n}{\vec{k}} \cdot 
   p\big(\vec{k}\big) \cdot q(n)
\\[+0.8em]
& = &
\displaystyle \sum_{n\in\NNO}\, e^{-K\lambda} \cdot 
   \frac{(K\lambda)^{n}}{n!} \cdot \left(\sum_{\vec{k}\in\som^{-1}(n)} 
   \frac{\binom{n}{\vec{k}}}{K^n} \cdot p\big(\vec{k}\big)\right) \cdot q(n)
\\[+0.8em]
& = &
\displaystyle \sum_{n\in\NNO} \; \poisson[K\lambda](n) \cdot 
   \big(\som^{\dag} \pull p\big)(n) \cdot q(n)
\\[+0.5em]
& = &
\big(\poisson[K-]\big)(\lambda) \models \big(\som^{\dag} \pull p\big) \andthen q.
\end{array} \]

\end{document}